\documentclass[final,5p,times,twocolumn,authoryear]{elsarticle}

\usepackage{amssymb}
\usepackage{amsmath}

\usepackage{lineno}
\usepackage[inline]{enumitem}
\usepackage{amsthm}
\usepackage{thmtools}
\usepackage{subcaption}
\usepackage{xurl}
\declaretheoremstyle[%
  headfont=\bfseries,%
  headpunct={:},%
  notefont=\normalfont\bfseries,%
  notebraces={--~}{},%
  qed=$\blacksquare$,%
]{definitionstyle}

\declaretheorem[style=definitionstyle,name=Definition]{defn}

\theoremstyle{definition}

\theoremstyle{remark}

\usepackage{amsmath} 
\usepackage{nicematrix}
\NiceMatrixOptions{
code-for-first-row = \color{blue} ,
code-for-last-row = \color{blue} ,
code-for-first-col = \color{blue} ,
code-for-last-col = \color{blue}
}
\usepackage{hyphenat}
\usepackage{units}
\usepackage{multirow}
\usepackage{booktabs}
\usepackage{graphics}
\usepackage{pdfpages}
\usepackage{xr}
\makeatletter
\newread\ELF@in
\def\ELF@nl{\newlabel}
\long\def\ELF@test#1#2#3#4\ELF@end{\def\ELF@a{#1}%
  \ifx\ELF@a\ELF@nl\global\@namedef{r@#2}{#3}\fi}
\def\ELF@loop{\read\ELF@in to\ELF@line
  \ifeof\ELF@in\else
    \expandafter\ELF@test\ELF@line\relax\relax\relax\ELF@end
    \expandafter\ELF@loop\fi}
\newcommand{\externallabelsfile}[1]{\openin\ELF@in=#1\relax
  \ifeof\ELF@in\PackageWarningNoLine{main}{Labels file #1 not found}%
  \else{\makeatletter\endlinechar=-1 \ELF@loop}\closein\ELF@in\fi}
\makeatother
\externallabelsfile{supplement.tex}

\usepackage{xcolor}
\newif\ifshowrev \showrevtrue
\definecolor{revblue}{RGB}{0,70,170}

\journal{Journal of Hydrology}

\showrevfalse

\begin{document}

\begin{frontmatter}



\title{A Hetero-functional Graph State Estimator for Watershed Systems: Application to the Chesapeake Bay}


\author[label1]{Megan S. Harris\fnref{fn1}}
\author[label1]{John C. Little}
\author[label2]{Amro M. Farid}

\fntext[fn1]{Present address: Department of Civil, Construction and Environmental Engineering, North Carolina State University, Raleigh, NC 27695, USA.}
\affiliation[label1]{organization={Department of Civil and Environmental Engineering},
            addressline={Virginia Tech},
            city={Blacksburg},
            postcode={24061},
            state={Virginia},
            country={United States}}

\affiliation[label2]{organization={School of Systems and Enterprises},
            addressline={Stevens Institute of Technology},
            city={Hoboken},
            postcode={07030},
            state={New Jersey},
            country={United States}}

\begin{abstract}

Regional watersheds are systems of systems in which hydrology, land use, and governance interact. Nutrient models such as the Chesapeake Assessment Scenario Tool (CAST) route nutrients from land to the estuary through lumped delivery factors, so the river network that connects sources to the estuary is not represented explicitly. This paper develops a region-independent watershed reference architecture in Hetero-functional Graph Theory (HFGT) and adapts the Weighted Least Squares Error Hetero-functional Graph State Estimator to estimate nutrient flows in that network from uncertain data. The framework is applied to the Chesapeake Bay watershed. A mass-conserving network of land segments, outlet points, and river reaches is built from public geospatial data. CAST applied loads, land-segment edge-of-river loads, and river-to-bay delivery factors enter the estimator as measurements, each with its own error term. CAST-reported loads delivered to the Bay are withheld from estimation and used only to test the estimated delivery. The estimator reproduces the applied loads ($R^2=1.00$) and the land-segment edge-of-river loads ($R^2=0.997$) and under-predicts delivery to the Bay by 8.0\% for nitrogen and 5.7\% for phosphorus. The gap is traced to CAST delivery factors that imply in-river gains, which a mass-conserving network cannot represent, and to land segments whose edge-of-river loads exceed their applied loads. The framework provides a common systems representation for coupling watershed models with infrastructure, economic, and governance models.
\end{abstract}

\begin{graphicalabstract}
\includegraphics[width=\textwidth]{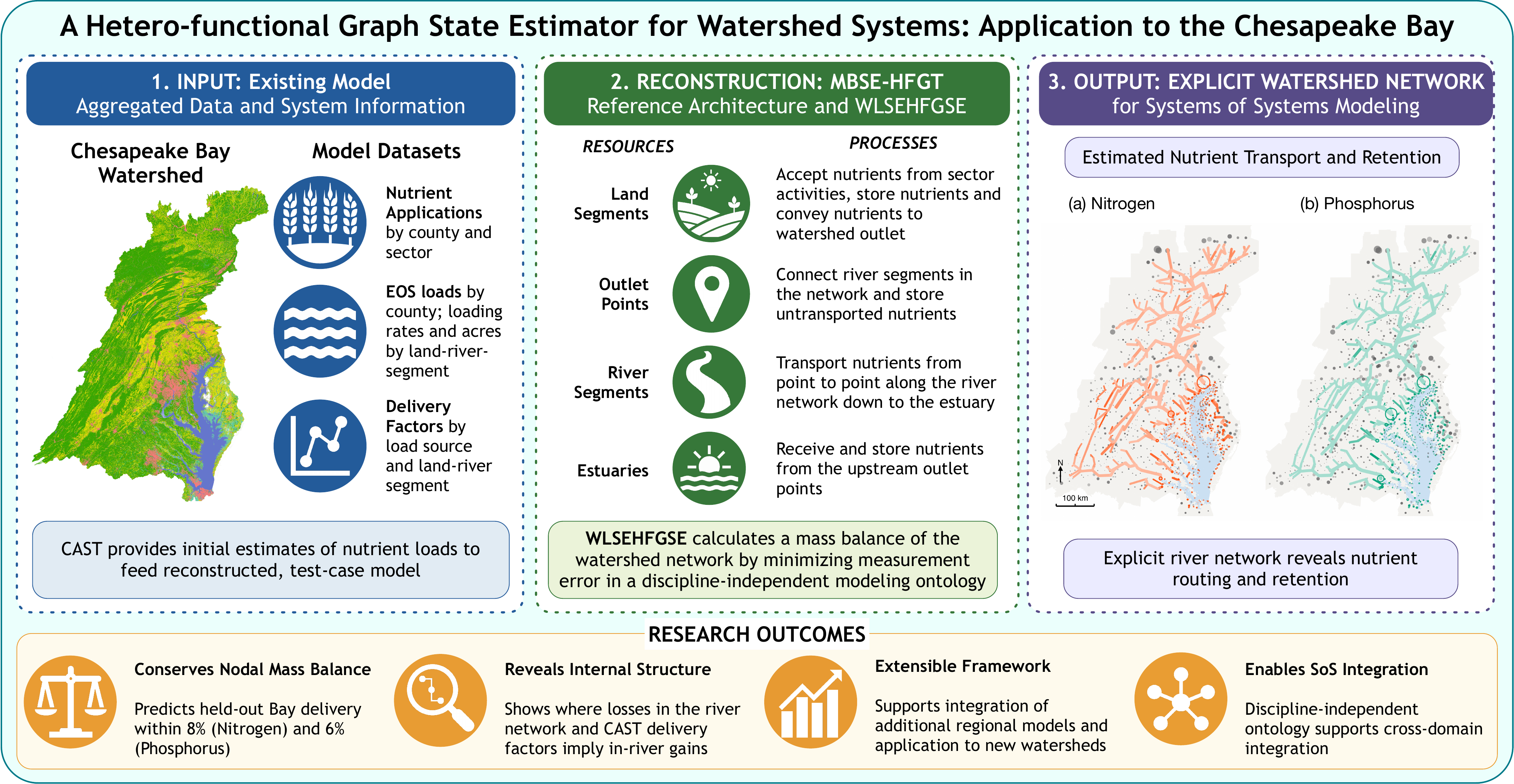}
\end{graphicalabstract}

\begin{highlights}
\item Offers a common ontology for coupling watersheds with economic and governance models
\item Watershed reference architecture defined in hetero-functional graph theory
\item Applies method to the Chesapeake Bay watershed using geospatial and behavioral data
\item State estimator infers nutrient flows in an explicit, mass-conserving river network
\item Bay delivery withheld from fitting is predicted within 8\% (N) and 6\% (P)
\end{highlights}

\begin{keyword}
watershed modeling \sep
Model-Based Systems Engineering (MBSE) \sep
Hetero-functional Graph Theory (HFGT) \sep
system of systems \sep
Systems Modeling Language (SysML) \sep
nutrient transport
\end{keyword}

\end{frontmatter}


\section{Introduction} \label{sec:cbw-introduction}
The modern societal challenges reflected in the Sustainable Development Goals (SDGs) are inherently interconnected, arising from interactions between multiple systems of the Earth and society known as \textit{systems of systems} (Defn.~\ref{defn:cbw-SoS}) ~~\citep{Folke:2021:00,rockstrom:2015:00,biermann:2016:00,Little:2023:ISC-J50}. These challenges depend on dynamic feedbacks between human and natural processes
\citep{Biermann:2022:00,bodin:2017:00,reed:2008:00}. Addressing them requires not only understanding each system individually, but also their interdependencies and how flows of water, energy, nutrients, and information move across disciplinary and spatial boundaries. 

\begin{defn}[System of Systems]\label{defn:cbw-SoS}
1) System of interest whose system elements are themselves systems; typically, these entail large-scale interdisciplinary problems with multiple, heterogeneous, distributed systems~\citep{SE-Handbook-Working-Group:2015:00}. 2) Set of systems or system elements that interact to provide a unique capability that none of the constituent systems can accomplish on its own~\citep{ISO-IEC-IEEE:2019:00}.
\end{defn}

Motivating this manuscript, many contemporary societal challenges are rooted in environmental systems, including watershed management, nutrient pollution, land-use change, and climate adaptation. These challenges arise from interactions among hydrologic, ecological, atmospheric, and human systems operating across multiple spatial and temporal scales~\citep{liu:2007:01,An:2012:00, steffen:2018:00}.
Existing approaches for studying such systems generally fall into two categories: detailed domain-specific simulation or co-simulation frameworks~\citep{fatichi:2016:00,jonsson:2024:00} and aggregate data-driven approaches such as system dynamics~\citep{naderi:2021:00,An:2012:00,liu:2007:02,zhao:2018:01}. While simulation frameworks can capture complex system behavior in considerable detail, they are often developed within disciplinary boundaries and can be difficult to integrate across domains~\citep{Naderi:2026:ISC-JR08}. Conversely, aggregate approaches can facilitate analysis of broad system interactions but frequently do so at the expense of process fidelity and mechanistic representation~\citep{Naderi:2026:ISC-JR08}. This creates the need for a common modeling framework capable of representing heterogeneous systems under a shared formalism while preserving the ability to describe complex interactions across disciplines and scales~\citep{Little:2023:ISC-J50,Little:2026:ISC-JR10}.

Model-Based Systems Engineering (MBSE) provides a formal approach for representing system form, function, and interactions through a common ontology, setting the basis for modeling systems of systems that span disciplines under a unified language~\citep{henderson:2021:00,ramos:2012:00}. Building on the fundamentals of MBSE, Hetero-functional Graph Theory (HFGT) utilizes graph theory with explicitly defined resource and function interactions, enabling representation of heterogeneous systems within a common architecture~\citep{Farid:2022:ISC-J49,Schoonenberg:2019:ISC-BK04,Farid:2025:ISC-JR06}. 

A framework based on MBSE and HFGT has previously been applied to multi-modal and integrated energy networks~\citep{Thompson:2023:ISC-J51, Thompson:2024:ISC-J53, Thompson:2025:ISC-JR11, Farid:2022:ISC-AP80}, the structural resilience of electric power systems~\citep{Thompson:2021:SPG-J44, Thompson:2020:SPG-C68}, interdependent smart city infrastructures~\citep{Schoonenberg:2017:ISC-C60, Schoonenberg:2018:ISC-AP35}, personalized healthcare delivery systems~\citep{Khayal:2017:ISC-J35, Khayal:2021:ISC-J46}, economic input--output systems~\citep{naderi:2026:01} and reconfigurable manufacturing supply networks~\citep{Kelepouris:2006:IEM-C04, Schoonenberg:2017:IEM-J34}.
Recent work by Harris et al.~\citep{harris:2026:io} demonstrated the feasibility of representing hydrologic systems using MBSE and HFGT. However, the study focused on illustrative examples and did not employ the use of a state estimator to apply the framework to a real-world watershed system with uncertain data.
Complementarily, Thompson and Farid~\citep{Thompson:2025:ISC-JR11} developed the Weighted Least Squares Error Hetero-functional Graph State Estimator (WLSEHFGSE), a generalized optimization and estimation framework for integrating heterogeneous and partially observed data. Watershed systems are characterized by sparse observations, uncertain measurements, and heterogeneous data sources, making state estimation a critical challenge.

While these developments establish the methodological foundation, applying MBSE-HFGT to real-world watershed systems also requires consideration of the modeling frameworks currently used in hydrologic practice.
Existing watershed modeling efforts have produced a diverse collection of hydrologic and water-quality models that differ in their governing assumptions, spatial and temporal resolutions, and intended applications~\citep{devia:2015:00,keller:2023:00}. In the United States, many regions have relied on the process-based Hydrological Simulation Program--FORTRAN (HSPF) to represent watershed hydrology and nutrient transport~\citep{keller:2023:00,bicknell:1997:00}. These regional watershed models are often adapted or extended. For example, the Chesapeake Bay Program developed a planning and assessment platform, the Chesapeake Assessment Scenario Tool (CAST) that emulates the behavior of the underlying HSPF-based watershed modeling framework for scenario evaluation~\citep{chesapeake-bay-program:2020:00}. Likewise, the Soil and Water Assessment Tool (SWAT) and its successor SWAT+ are widely applied for watershed-scale analyses of streamflow, sediment transport, nutrient dynamics, and land-management impacts across agricultural and mixed-use landscapes~\citep{yonce:2019:00,yamini-priya:2021:00}. Other frameworks, including integrated groundwater--surface water models such as APEX-MODFLOW, have been developed to address specific hydrologic processes and regional management needs~\citep{bailey:2021:00}.

Although these models differ in implementation and scope, they share common structural concepts, including the transport and transformation of water, sediment, nutrients, and other environmental constituents across interconnected watershed components. Consequently, a structural modeling framework that is independent of any particular watershed model has the potential to provide a common representation through which diverse regional modeling approaches can be integrated and analyzed.

These advances establish a foundation for applying MBSE and HFGT to real-world, multi-scale environmental systems for future system of systems modeling across disciplines.
Many of these regional models, including the Phase 6 Chesapeake Bay watershed model, use statistically fitted delivery factors rather than an explicit river network to represent land-to-estuary transport~\citep{Hood:2021:00}. These factors estimate how much mass reaches the estuary but not its transport path, where losses occur, or whether delivery is consistent with network-wide mass conservation.
This paper therefore asks: can a lumped regional nutrient model be re-expressed as an explicit, mass-conserving network within a system-independent ontology, and what does that network reveal about the model's delivery structure?

\subsection{Original Contribution}


This paper presents a modeling framework for representing watershed systems and demonstrates its application to the Chesapeake Bay watershed. The framework integrates open-source geospatial processing with a state estimation methodology grounded in HFGT. First, a region-independent watershed reference architecture is developed within the HFGT meta-architecture. Watershed dynamics are then formulated using the Weighted Least Squares Error Hetero-functional Graph State Estimator~\citep{Thompson:2025:ISC-JR11}, which extends the Hetero-functional Network Minimum Cost Flow formulation~\citep{Farid:2022:01} by incorporating estimation error into the measurement equations. Each data source enters as its own measurement with its own error term, so unknown or partially observed nutrient flows are inferred from data that may be mutually inconsistent while mass is conserved exactly at every node.

The framework is demonstrated using the Chesapeake Assessment Scenario Tool (CAST) of the Chesapeake Bay Program. A reproducible Python pipeline constructs hetero-functional incidence tensors from publicly available geospatial datasets, including digital elevation models, stream networks, river segment polygons, and estuarine boundaries. The resulting network explicitly identifies outlet points, resolves routing connections between river segments, and represents watershed topology in a transparent form that is consistent with the assumptions of CAST. CAST applied loads, land-segment edge-of-river loads, and river-to-bay delivery factors then enter the estimator as measurements. CAST's total load delivered to the Bay is withheld from estimation and used to test whether the network preserves CAST's delivery when its inputs are imposed as uncertain measurements.

The primary contributions of this work are:
\begin{enumerate*}[label=\arabic*),labelsep=1em]
\item a transferable reference architecture for watershed systems within the HFGT meta-architecture,
\item a watershed state estimation methodology based on the WLSEHFGSE in which each data source has its own measurement error,
\item an application to the Chesapeake Bay watershed in which CAST's delivered load to the Bay is withheld from estimation and predicted,
\item a diagnosis, made possible by the explicit network, of where CAST's fitted delivery factors cannot be reproduced by mass-conserving routing, and
\item a reproducible workflow for integrating heterogeneous geospatial and watershed model data into a common systems representation.
\end{enumerate*}

Together, these contributions establish a common representation for watershed systems that can be extended to other regions and integrated with existing HFGT models of infrastructure, energy, governance, and other interacting systems.

\subsection{Paper Outline}

The remainder of the paper is organized as follows. Section~\ref{sec:cbw-testCase} introduces the Chesapeake Bay watershed as the chosen region of study for the test case.
Section~\ref{sec:cbw-methods} presents the methodological framework, and 
Section~\ref{sec:cbw-applicationCBW} applies the proposed methodology to the Chesapeake Bay watershed. 
Section~\ref{sec:cbw-results} presents the resulting watershed representation and state-estimation results and discusses their implications. Finally, Section~\ref{sec:cbw-conclusion} summarizes the principal contributions of the paper and identifies opportunities for future research.

For readers seeking additional background, the Supporting Material provides a comprehensive overview of the HFGT meta-architecture and the WLSEHFGSE framework. 

\section{Test Case: Chesapeake Bay Watershed} \label{sec:cbw-testCase}
This paper utilizes the Chesapeake Bay watershed (CBW) as a test case towards integrating watershed systems of systems into the MBSE-HFGT methodology.  
The CBW region is a system of systems spanning hydrology, estuarine ecology, agriculture, infrastructure, regional economics, and policy.
The region is an ideal testbed for translating the MBSE–HFGT framework from illustrative systems to real-world applications based on its size, data availability, complexity, and economic and policy relevance. 
\subsection{Region of Study} \label{subsec:cbw-regionOfStudy}
The Chesapeake Bay watershed spans six states and the District of Columbia and includes the largest estuary in the United States~\citep{chesapeake-bay-program:2020:02}. The estuary itself, the final destination of watershed-routed water and nutrients, is one of the most comprehensively modeled and monitored ecosystems in the world~\citep{Hood:2021:00,usepa-u.s.-environmental-protection-agency:2010:00,easton:2017:00}.
Physiographic diversity from the Appalachians through the Piedmont to the Coastal Plain creates strong gradients in runoff generation, transport, and attenuation~\citep{usepa-u.s.-environmental-protection-agency:2010:00}. 

These interacting processes contribute to long-standing environmental degradation, including eutrophication and hypoxia~\citep{hagy:2004:00,kemp:2005:00}, and long-term declines in water quality~\citep{williams:2010:00}. This degradation impacts fishery productivity~\citep{buchheister:2013:00}, public health~\citep{landsberg:2006:00}, and water clarity valued by residents and tourists~\citep{klemick:2018:00}. With ecosystem services valued at over \$100 billion annually~\citep{chesapeake-bay-foundation:2014:00}, understanding the coupled dynamics of the CBW remains a regional and national priority.

\subsection{The Chesapeake Bay Modeling Suite}

The Chesapeake Bay Program operates a highly integrated modeling suite that connects watershed, airshed, and estuary dynamics. Over years of development, the watershed model has shifted forms.

From Phase 1 (1985) through Phase 5 (2010), the CBP watershed model was mechanistic and based in HSPF~\citep{usepa-u.s.-environmental-protection-agency:2010:00}. These versions of the model ran at hourly intervals to simulate hydrology, sediment, and nutrient cycling across segments, with extensive calibration and linkage to a 3-D estuarine water-quality and sediment-transport model (WQSTM/CH3D-ICM)~\citep{usepa-u.s.-environmental-protection-agency:2010:00}. Foundational documentation spans the HSPF manuals and CBP watershed applications~\citep{bicknell:1997:00,chesapeake-bay-program:1998:00}, with airshed inputs provided via CMAQ/regression frameworks and cross-media coupling~\citep{linker:1999:00,Hood:2021:00}. This era preserved explicit routing and process couplings, where annual management metrics were produced by aggregating time-varying simulations.

Phase 6 (2017) created CAST, a time-averaged (annual) model that constrains the previous dynamic model to produce aggregated nitrogen, phosphorus, and sediment loads from the land to the estuary~\citep{Hood:2021:00}. 
Geospatially, CAST partitions the region into land-river segments which are the intersection of county and watershed boundaries. The model produces annual Edge of Stream (EOS) and End of Tide (EOT) delivered loads aggregated by county~\citep{chesapeake-bay-program:2020:00}. 
Computationally, CAST functions as a statistically calibrated emulator: average-load functions, best management practice (BMP) effects, and delivery factors (land to water, stream, river) are derived from mechanistic model outputs, literature syntheses, and empirical fits to monitoring data in a regression-style structure~\citep{Hood:2021:00}. For example, land-to-water and stream delivery were informed by USGS/SPARROW-style relations and empirical modeling~\citep{ator:2011:00,ator:2016:00}, while river-to-bay delivery factors were constrained using the dynamic watershed-estuary model~\citep{Hood:2021:00,cerco:2017:00}.

For each land-river segment and load source, CAST computes delivered load as a product: the edge-of-stream (EOS) load times a stream-to-river factor gives the edge-of-river (EOR) load, and the EOR load times a river-to-bay factor gives the end-of-tide (EOT) load delivered to the estuary~\citep{chesapeake-bay-program:2020:00}. All three factors are fitted quantities rather than measurements.

The CAST watershed model is publicly accessible for stakeholder use~\citep{chesapeake-bay-program:1998:00}. The Phase 6 model improved interpretability, computational efficiency, and integration among watershed, land-use, estuarine, and airshed modeling components~\citep{Hood:2021:00}. However, these improvements were accompanied by several simplifications in transitioning from the mechanistic watershed model to CAST. First, the internal routing of water was removed and nutrient transport was reduced to delivery factors that route nutrients from land to rivers and ultimately to the estuary, limiting transparent representation of hydrologic connectivity and processes~\citep{Hood:2021:00}.
Temporal resolution was also reduced from hourly dynamics to annual aggregations, limiting time-dependent simulation~\citep{Hood:2021:00}. Furthermore, interactions among the watershed, estuarine, airshed, and land-use models remain only loosely coupled~\citep{Hood:2021:00}. Consequently, CAST provides an efficient and accessible platform for stakeholder analysis but with limited ability to represent internal watershed dynamics and coupled system interactions.

These limitations become increasingly important as the Chesapeake Bay watershed region faces emerging environmental challenges that depend on interactions among multiple interconnected systems operating across spatial and temporal scales.
Understanding and managing such dynamics therefore requires modeling frameworks capable of representing cross-scale feedbacks and evolving data streams~\citep{lefcheck:2017:00,ni:2020:00,testa:2018:00,irby:2016:00,irby:2018:00,irby:2019:00}.




\subsection{Data Utilization}

The Chesapeake Bay Modeling Suite, specifically CAST, serves three roles in this study. First, CAST provides a starting point for the development of a watershed reference architecture within the MBSE-HFGT framework. Second, CAST informs the derivation of a watershed-system model and associated state-estimation formulation under the HFGT ontology. Third, CAST provides data for an initial demonstration of the proposed methodology using the Chesapeake Bay watershed as a test case.

While the watershed component of the Chesapeake Bay Modeling Suite is ultimately derived from a fully dynamic HSPF-based hydrologic and water-quality modeling framework~\citep{bicknell:1997:00,usepa-u.s.-environmental-protection-agency:2010:00}, this work deliberately focuses on CAST as the primary point of integration with MBSE-HFGT. CAST is the operational tool used by Chesapeake Bay Program stakeholders and is publicly accessible through a web-based interface. Consequently, its structure, assumptions, and outputs directly inform planning and management activities throughout the watershed. 
As such, this preliminary implementation is limited by the data available through CAST. Because CAST does not retain explicit water tracking, the present model is limited to nitrogen and phosphorus.

The selection of CAST was also informed by sustained engagement with Chesapeake Bay Program personnel through an NSF-supported research collaboration. These interactions
provided opportunities to discuss CAST objectives, structure, and assumptions.
Although the framework developed in this manuscript is demonstrated using CAST, the MBSE-HFGT methodology is model-agnostic and can provide a common structural representation through which multiple watershed models and region-specific modeling frameworks can be integrated within a broader co-modeling or system-of-systems architecture.

Future work can evaluate the resulting representation against monitoring data, the underlying HSPF-based dynamic model, and coupled hydrologic--economic frameworks built upon HSPF-type process models~\citep{Amaya:2021:00,Amaya:2022:00,Abdolabadi:2023:00}. Additionally, the framework could be extended to explicitly represent water movement throughout the watershed using information derived from HSPF and SWAT models~\citep{yonce:2019:00,devia:2015:00,yamini-priya:2021:00}. Incorporating these hydrologic processes would enable the tracking of nutrient concentrations, in addition to nutrient loads, across the region.

Importantly, the objective of this study is not to reproduce or replace the Chesapeake Bay Modeling Suite. Rather, CAST serves as an accessible and extensible foundation for developing a generalized watershed representation within the MBSE-HFGT framework.



\section{Methodology} \label{sec:cbw-methods}
This section provides the methodological foundation for the proposed framework for representing watershed systems using the MBSE-HFGT framework. Subsection~\ref{subsec:cbw-defineHFGTwatershed} defines the watershed-specific reference architecture under the HFGT meta-architecture, and Subsection~\ref{subsec:cbw-deriveHFGTwatershed} derives the mathematical model of this watershed system based on HFGT equations and data structures.

Readers new to HFGT are referred to the conceptual introduction by Farid et al.~\citep{Farid:2025:ISC-JR06}, and worked examples for simple hydrologic systems are given in~\citep{harris:2026:io}.

\subsection{Defining Watershed Systems within the HFGT Meta-Architecture} \label{subsec:cbw-defineHFGTwatershed}
This subsection applies the HFGT meta-architecture and definitions to watershed systems. These definitions are elaborated in the supplementary materials, Section~\ref{sec:cbw-hfgt}.

\begin{figure}[h!]
    \centering
    \includegraphics[width=\linewidth]{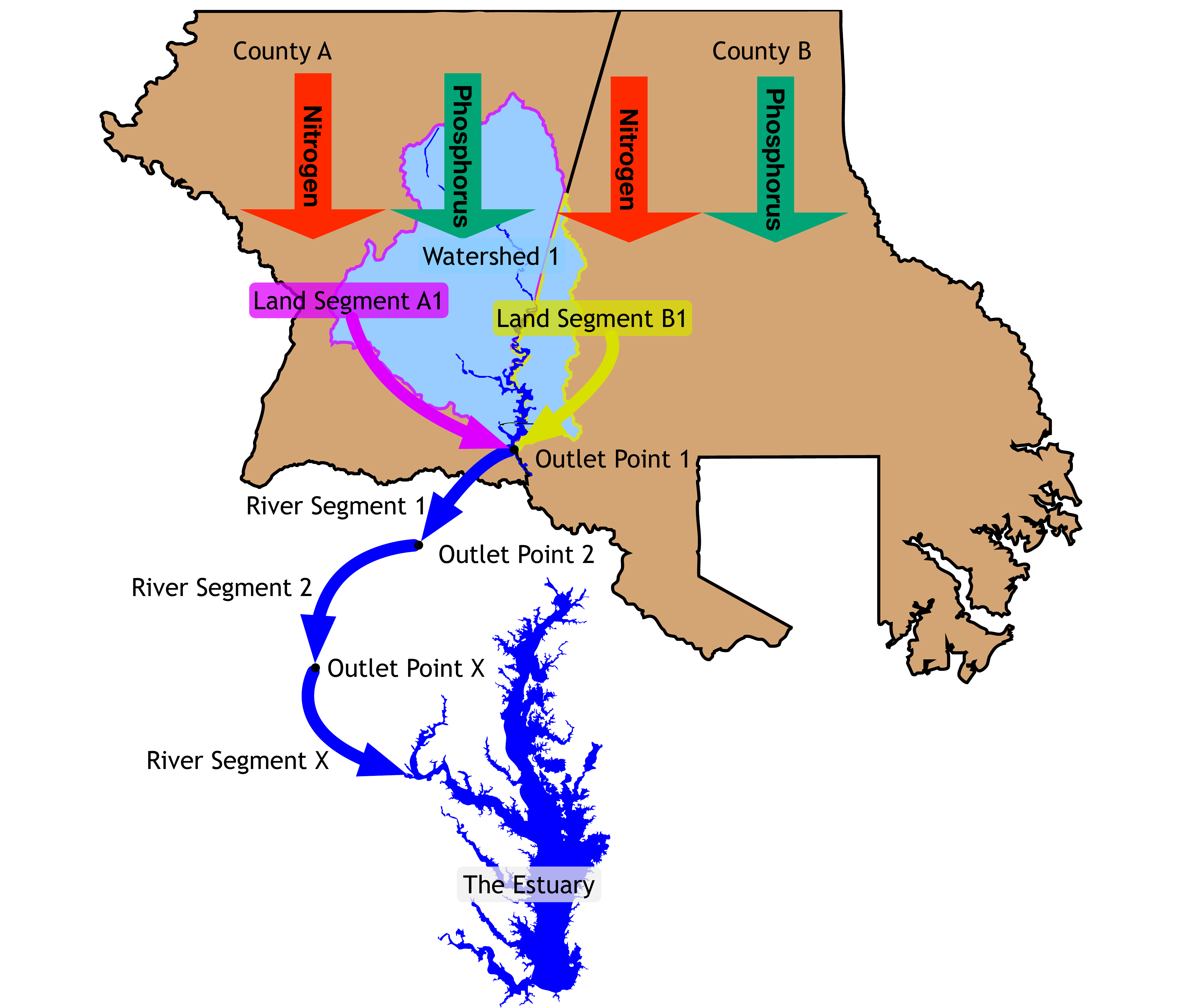}
    \caption{Schematic representation of nutrient flow across land segments, defined as the intersection of watershed and county boundaries. 
    The Liberty Dam Watershed is shown as an example, with Carroll County (A) and Baltimore County (B) highlighted \protect\citep{chesapeake-bay-program:2024:00}.}
    \label{fig:cbw-systemDiagram}
\end{figure}

Informed by CAST, the generic structure of watershed systems is visualized in Fig.~\ref{fig:cbw-systemDiagram}. These systems include land segments, outlet points, river segments, and the estuary. This generic structure helps readers visualize the systems thinking abstractions of the real-world system (i.e. the Chesapeake Bay watershed system) that informed the reference architecture developed in this manuscript. 

As visualized in Fig.~\ref{fig:cbw-systemDiagram}, the watershed system nutrient dynamics depend on three interacting domains: land use, watershed, and economic. On the economic side, nutrients are applied as fertilizer or manure for crop production, as a by product of animal husbandry, or industrial processes~\citep{chesapeake-bay-program:2020:00}. From a data perspective, these applications are reported at the county scale~\citep{chesapeake-bay-program:2020:00}. Land use characteristics then influence nutrient fate. For example, highly developed land facilitates runoff with limited infiltration, leading to higher nutrient concentrations in surface water. Finally, subwatersheds of varying size route water from any point on the land within their boundary to an outlet point in the river network~\citep{chesapeake-bay-program:2020:00}. This river network accumulates flows across increasing stream orders, ultimately discharging into the estuary~\citep{chesapeake-bay-program:2020:00}.

Based on this system abstraction, Paragraph \ref{subsubsec:cbw-bddwatershed} develops a watershed Block Definition Diagram (BDD) that is a specialization of the HFGT meta-architecture BDD shown in Fig.~\ref{fig:cbw-LFESMetaArchitecturebdd} of the Supporting Information, and Paragraph \ref{subsubsec:cbw-actwatershed} develops an Activity Diagram (ACT) that is a specialization of the HFGT meta-architecture ACT shown in Fig.~\ref{fig:cbw-LFESMetaArchitectureact} of the Supporting Information. These graphical depictions together make up a watershed Reference Architecture (RA) that allows the application in Paragraph \ref{subsubsec:cbw-hfgtwatersheddefn} of the definitions from Supporting Information Section~\ref{sec:cbw-hfgt}. The goal is to represent the application and transport of nutrients, specifically nitrogen and phosphorus, which strongly influence aquatic ecosystem health. 

\subsubsection{Block Definition Diagram for a Generic Watershed System} \label{subsubsec:cbw-bddwatershed}

\begin{figure*}[hbt!]
    \centering
    \includegraphics[width=\linewidth]{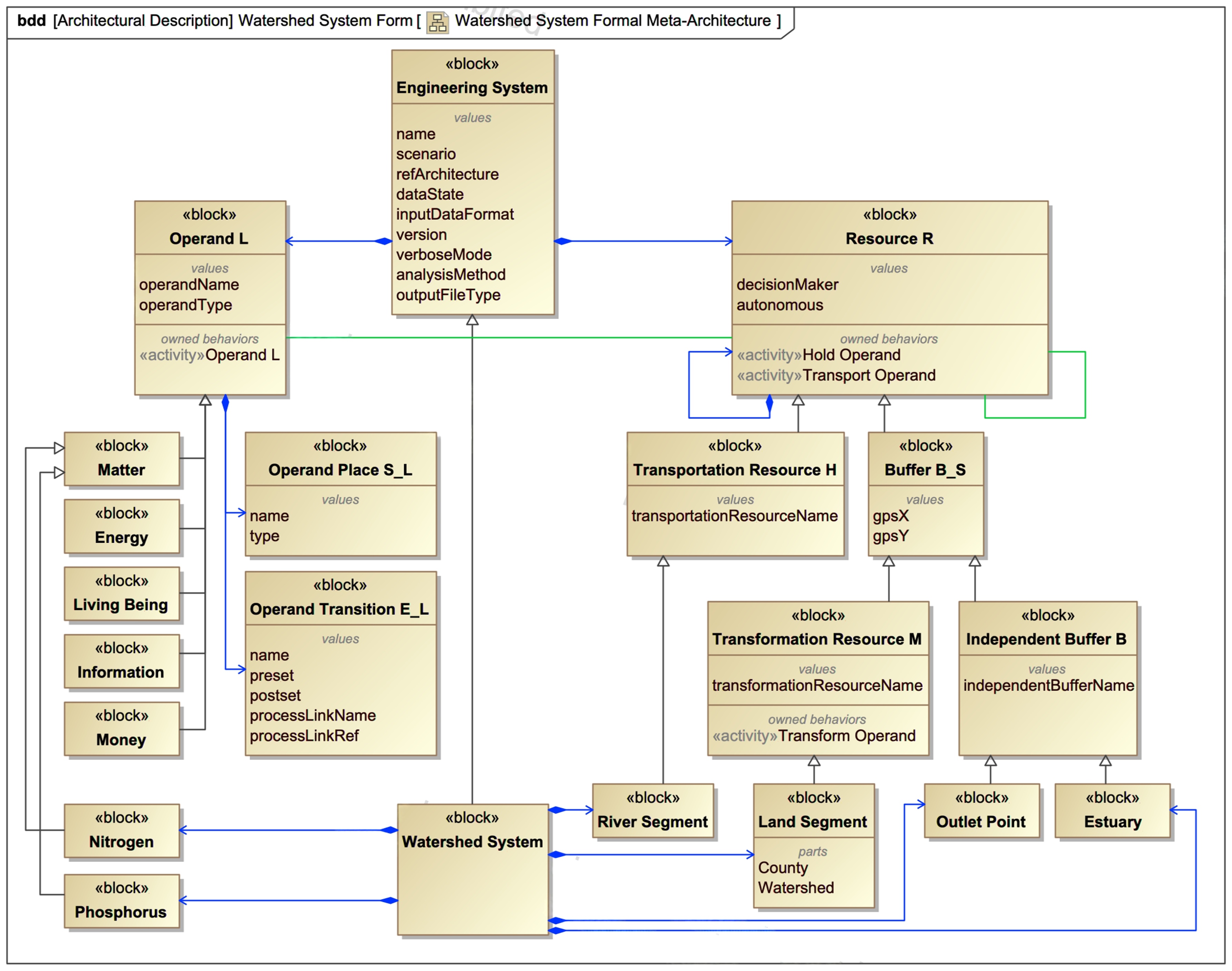}
    \caption{Block Definition Diagram of a Generic Watershed System}
    \label{fig:cbw-bdd-watershed}
\end{figure*}
This paper establishes watershed systems as a distinct class of engineering systems by defining a watershed BDD (Fig.~\ref{fig:cbw-bdd-watershed}) under the engineering system meta-architecture BDD (Fig.~\ref{fig:cbw-LFESMetaArchitecturebdd}). This watershed system is a class of engineering system whose form is summarized to include the following:
\begin{enumerate*}[label=\arabic*),labelsep=1em]
    \item land segments, represented by polygons defined by unique combinations of county and river segment,
    \item outlet points, where all land segments within a watershed drain,
    \item river segments, which link outlet points to their next downstream outlet point,
    \item estuaries, where watershed flows discharge, and
    \item nutrients, nitrogen and phosphorus, which flow throughout the system.
\end{enumerate*} 

Within the watershed system, nitrogen and phosphorus serve as the primary quantities being transported and transformed and are therefore treated as \textit{operands} (Defn.~\ref{def:cbw-operand}) in the HFGT framework. The physical components of the watershed that facilitate these processes---including land segments, river segments, outlet points, and estuaries---are represented as \textit{resources} (Defn.~\ref{def:cbw-resource}).

Resources can be categorized according to their role in the movement and storage of operands. Land segments act as \textit{transformation resources}, where nutrients may be stored, transported, and introduced into the system (considered a transformation process in HFGT). Outlet points and estuaries serve as \textit{independent buffers}, representing locations where nutrients may be retained.
Collectively, land segments, outlet points, and estuaries are classified as \textit{buffers} (Defn.~\ref{def:cbw-buffer}) because they provide storage for operands, in this case nutrients, as they move through the watershed system. Finally, river segments function as \textit{transportation resources}, moving nutrients through the watershed.

\subsubsection{Activity Diagram of Watershed Systems}\label{subsubsec:cbw-actwatershed}
\begin{figure*}
    \centering
    \includegraphics[width=\linewidth]{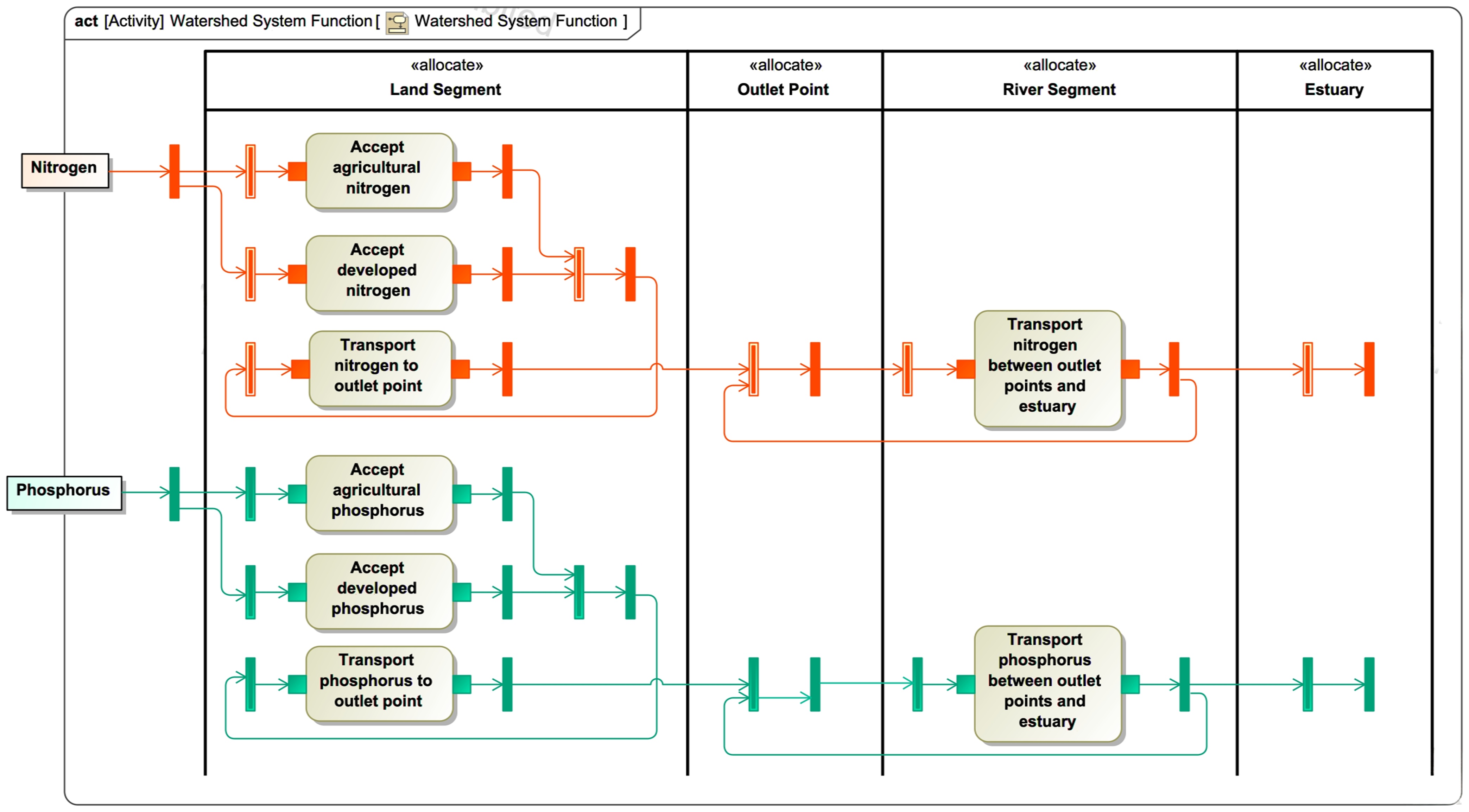}
    \caption{Activity Diagram of a Generic Watershed System}
    \label{fig:cbw-act-watershed}
\end{figure*}

The processes (Defn. \ref{def:cbw-process}) enabled by these resources are shown in the watershed-specific ACT (Fig.~\ref{fig:cbw-act-watershed}), which specializes the ACT defined by the engineering system meta-architecture (Fig.~\ref{fig:cbw-LFESMetaArchitectureact}). These processes transform and transport the system operands, nitrogen and phosphorus, as demonstrated in the HFGT meta-architecture formulation in Fig.~\ref{fig:cbw-LFESMetaArchitectureact} of the Supporting Information. 

Each resource class possesses a set of \textit{capabilities} (Defn.~\ref{def:cbw-capability}), which define the processes that may occur at that resource. Land segments can, 
\begin{enumerate*}[label=\arabic*),labelsep=1em]
    \item accept nitrogen from agricultural land (accept agricultural nitrogen),  
        \item accept phosphorus from agricultural land (accept agricultural phosphorus),  
        \item accept nitrogen from developed land (accept developed nitrogen),  
        \item accept phosphorus from developed land (accept developed phosphorus),
        \item transport nitrogen from the land segment to its outlet point, and
        \item transport phosphorus from the land segment to its outlet point
\end{enumerate*}, where agricultural and developed nutrients refer to the sector applying the nutrient to the land.
River segments can 
\begin{enumerate*}[label=\arabic*),labelsep=1em]
    \item transport nitrogen from outlet point to outlet point or to the estuary and
    \item transport phosphorus from outlet point to outlet point or to the estuary.
\end{enumerate*}
The formulation enforces explicit conservation of nutrient mass: whatever enters a land segment or outlet point and is not transported onward remains in that buffer's stock $Q_B$. Because the reference architecture has no separate loss processes (e.g., crop uptake, denitrification, or burial), $Q_B$ is interpreted throughout this paper as \emph{retention}, the mass that is not passed downstream within the time step, and not as physical accumulation. Retention in a land segment is the applied mass that does not reach the river network; retention at an outlet point is the in-river loss on the reach that begins there. In the ACT (Fig.~\ref{fig:cbw-act-watershed}), these dynamics appear as the vertical joins and forks linking operand flows within each buffer class: land segment, outlet point, and estuary. In this way, the model infers how nitrogen and phosphorus that are applied upstream are released at downstream locations in the watershed.

\subsubsection{HFGT Definitions}\label{subsubsec:cbw-hfgtwatersheddefn}


With the RA for the generic watershed system defined, the watershed system is classified by the HFGT definitions introduced in the Supporting Information Section~\ref{sec:cbw-hfgt}. These terms form the following sets and indices, adopting the tensor-based notation of HFGT introduced in Farid et al.~\citep{farid2022tensor}:

\begin{trivlist}
\item \textbf{Operands $l_i \in L$ (Defn.~\ref{def:cbw-operand}):} In this watershed system, $L = \{N, P\}$.  Notably, water is not included amongst the system's operands because the CAST software does not track water flows. This modeling demonstrates how the WLSEHFGSE is not just applicable to different systems but can also be tailored to accommodate simplifying modeling assumptions about those systems. Future extension could easily incorporate water data to be able to infer nutrient concentrations throughout the system rather than simply nutrient masses.  
\item \textbf{Processes $p_w \in P$ (Defn.~\ref{def:cbw-process}):} represent actions operating on operands. As shown in Fig.~\ref{fig:cbw-act-watershed}, these include transformation processes that accept agricultural nitrogen, developed nitrogen, agricultural phosphorus, and developed phosphorus on land segments, and transportation processes that move the nitrogen and phosphorus from land segments to outlet points, to river segments.  As shown in Fig.~\ref{fig:cbw-act-watershed}, these processes are realized by land segments and river segments.  

\item \textbf{Resources $r_v \in R$ (Defn.~\ref{def:cbw-resource}):}
\begin{itemize}[noitemsep]
\item Each land segment is a transformation resource $m \in M \subset R$.   The boundaries of each land segment are defined by a polygon formed by the intersection of a county and a river segment.  
\item Each river segment is a transportation resource $h \in H \subset R$.  
\item Each outlet point or estuary is an independent buffer $b \in B \subset R$. 
\end{itemize}
\item \textbf{Buffers $b_{sy} \in B_S = M \cup B $ (Defn.~\ref{def:cbw-buffer})} :   include land segments, outlet points and estuaries.  
\item \textbf{Capabilities $\epsilon_\psi \in {\cal E}_S$ (Defn.~\ref{def:cbw-capability}):} Fig.~\ref{fig:cbw-act-watershed} shows 8 types of capabilities:
\begin{itemize}[noitemsep]
\item Land segment accepts agricultural nitrogen.
\item Land segment accepts developed nitrogen.
\item Land segment transports nitrogen to outlet point.
\item Land segment accepts agricultural phosphorus.
\item Land segment accepts developed phosphorus.
\item Land segment transports phosphorus to outlet point.
\item River segment transports nitrogen between outlet points and estuary.  
\item River segment transports phosphorus between outlet points and estuary.  
\end{itemize}
These are instantiated with each instantiated resource in the watershed.  
\end{trivlist}

\subsection{Deriving a Watershed System State Estimation Model} \label{subsec:cbw-deriveHFGTwatershed}
This subsection applies the generic WLSEHFGSE, as derived in the Supporting Information Section~\ref{sec:cbw-WLSEHFGSE}.  This model estimates operand flows over time incorporating exogenous data from specific watershed, land use and economic systems while preserving physical and structural constraints. Importantly, the WLSEHFGSE is stated generically, enabling its application to a variety of complex systems of systems. Specific systems, however, place limiting conditions on this general form.  When the general form of the WLSEHFGSE 
incorporates the physical characteristics of watershed systems, it collapses to the following specialized form:  

\begin{subequations}\label{eq:cbw-WLSEHFGSE-Water}
\begin{align}
\min\quad
Z
&=
\sum_{k=1}^{K-1}
\Big[
{\cal E}_{U}^T[k]F_{\cal E}{\cal E}_{U}[k]
+ U^{T}[k]A_{U}U[k]
\nonumber\\
&\qquad\qquad
+ Q_{B}[k]^TA_{Q_B}Q_{B}[k]
\Big]
\label{eq:cbw-WLSE-watershed-Obj}
\end{align}
subject to
\begin{align}
-Q_{B}[k+1] + Q_{B}[k]
+ (M^+-M^-)U[k]\Delta T
&=0
&& \forall k
\label{eq:cbw-WLSE-watershed-st}
\\
D_{U}U[k] - C_{U}[k]
&={\cal E}_{U}[k]
&& \forall k
\label{eq:cbw-WLSE-watershed-exogenous}
\\
Q_{B}[1]
&=0
\label{eq:cbw-WLSE-watershed-initCond}
\end{align}

where the decision variables are

\begin{align}
x[k]
=
\begin{bmatrix}
Q_B\\
U\\
{\cal E}_U
\end{bmatrix}[k],
\qquad
\forall k\in\{1,\dots,K\}.
\label{eq:cbw-WLSEHFGSE-Decision-Variables}
\end{align}
\end{subequations}
to account for buffer stocks of each operand $Q_{B}[k]$, the flows of operands through capabilities $U[k]$, and the measurement error ${\cal E}_{U}[k]$ associated with exogenous data $C_{U}[k]$.  

\subsubsection{Derivation of the Specialization}


The watershed-specific form of the WLSEHFGSE stated in Eq.~\ref{eq:cbw-WLSEHFGSE-Water} is derived from the general form in Eq.~\ref{eq:cbw-WLSEHFGSE} by incorporating the following physical characteristics of watershed systems.  
\begin{itemize}[noitemsep]
    \item The objective function in Eq.~\ref{eq:cbw-WLSEHFGSEOF} simplifies to Eq.~\ref{eq:cbw-WLSE-watershed-Obj} when:
    \begin{itemize}[noitemsep]
        \item Only the dependence on the decision variables in Eq.~\ref{eq:cbw-WLSEHFGSE-Decision-Variables} is included, and the linear terms are eliminated.  
        \item The time-invariant, diagonal, quadratic coefficient matrix $F_{\cal E}$ imposes a weight on the square of each measurement error. This is expressed as in ${\cal E}_U^T[k]F_{\cal E}{\cal E}_U[k]$ in matrix algebra. 
        \item The time-invariant, diagonal, quadratic coefficient matrices $A_U$ and $A_{QB}$ impose small quadratic penalties to flows and buffers.  Values of $\alpha = 10^{-10}$ and $\beta = 10^{-12}$ are used for each of the diagonal terms in the respective matrices.  Mathematically, these penalty terms help the optimization solver identify a single, stable solution. 
        More specifically, the presence of lumped, aggregated data can cause equifinality~\citep{beven:1992:00}. These quadratic penalty terms differentiate between equifinal solutions and encourage the solver to find the solution that distributes, rather than centralizes, nutrient stocks and flows. Without these small quadratic penalties, the problem has many equally optimal solutions, and the solver may return one in which flow does not propagate from the land segments to the estuary.
    \end{itemize}
    \item $k_{d\psi} = 0$ $\forall k \in \{1,...K\}$, $\forall \psi \in \{1,...,{\cal E}_S\}$: This RA assumes data inputs for all capabilities represent annual totals that occur instantaneously within the temporal resolution of the chosen annual time step. As a result,
    \begin{itemize}[noitemsep]
        \item Eq.~\ref{eq:cbw-EqualityConstraintsOC} simplifies to \ref{eq:cbw-WLSE-watershed-st} to describe the mass balance of nitrogen and phosphorus throughout the watershed. Nutrient tracking requires explicit buffer stocks, $Q_B$.  Each land segment, outlet, and the estuary is represented as a buffer with its own mass balance.   
        \item Eq.~\ref{eq:cbw-TransitionConstraint2} simplifies to $Q_{\cal E}[k+1] - Q_{\cal E}[k] = 0$ $\forall k \in {1,\ldots, K}$ and can therefore be removed.  
        \item Eq.~\ref{eq:cbw-DurationConstraint2} simplifies to $U^{+}[k] = U^{-}[k] = U[k]$ and can therefore be removed.  A new variable, $U[k]$ is introduced to simplify notation. 
        \item Eq.~\ref{eq:cbw-ESNMeasurement} simplifies to \ref{eq:cbw-WLSE-watershed-exogenous} to describe the exogenously defined firing vector weights, or flows of operands nitrogen and phosphorus, of the watershed system.
    \end{itemize}
    \item $S_{l_i}=0$ $\forall l_i \in L$: This watershed system assumes that the state of nitrogen and phosphorus do not change. All nitrogen and phosphorus are counted as total mass inclusive of all chemical formulations. Therefore, there are no operand nets that track the state of these operands. As a result, Eqs. \ref{eq:cbw-SSN-STF-QB2}-\ref{eq:cbw-SyncMinusOC} and \ref{eq:cbw-SSNMeasurement} are removed.  
    \item Eq.~\ref{eq:cbw-HFGTprog:comp:InitOC} simplifies to \ref{eq:cbw-WLSE-watershed-initCond}, setting the initial value of retained nutrients to zero at the beginning of the simulation.
    \item For Eq.~\ref{eq:cbw-QPcanonicalform:3OC}, the absence of data constraining the upper and lower bounds of nutrient flows means that this constraint can be removed.  
    \item For Eq.~\ref{eq:cbw-wlsehfgse-deviceModels}, the CAST software does not include effort variables $Y$ (e.g., nutrient concentrations) because it does not track water flows.  Consequently, this constraint can be removed as well. However, future formulations that reinstate water as an operand would add nutrient concentration tracking here. 
\end{itemize}

\subsubsection{Weighted Least Squares Error Objective Function}

The objective function of the watershed system takes the form described in Eq.~\ref{eq:cbw-WLSE-watershed-Obj}, minimizing the weighted sum of squared measurement errors ${\cal E}_U$, one for each measurement equation in Eq.~\ref{eq:cbw-WLSE-watershed-exogenous}.
CAST does not report uncertainties for the quantities used here, so this paper weights all measurement errors equally:
\begin{align}
F_{\cal E}= \tfrac{1}{2}\,I
\label{eq:cbw-weight-formula}
\end{align}
The factor $\tfrac{1}{2}$ is a scaling convention; because all errors share the same weight, it does not affect the solution apart from the relative size of the small regularization terms $A_U$ and $A_{Q_B}$. The objective is therefore the sum of squared errors in pounds, and measurements of large flows (large counties, main-stem reaches) carry more weight in absolute terms than measurements of small ones. Weighting each error by the inverse square of its measured value, $F_{\cal E}=\text{diag}\left(1/\max(C_U^2,2)\right)$, was also tested. However, because the CAST quantities span many orders of magnitude, the resulting weights made the problem numerically ill-conditioned and the solver did not reach a certified optimum. Equal weights also keep the 
results in Section~\ref{sec:cbw-results} independent of any tuned weighting. Weights based on data provenance and reliability are left to future work.

\subsubsection{Watershed Engineering System Net State Transition Function} \label{subsec:cbw-watershedTransitionFunction}
The state transition function (Eq.~\ref{eq:cbw-WLSE-watershed-st}) retains the full continuity equation with incidence matrix $M^\pm$. The incidence tensor in matrix form represents extraction of each nutrient-operand from buffers by capabilities and their injection back into new buffers.  Subsection~\ref{subsec:cbw-CASTtransitionFunction} outlines the construction of the hetero-functional incidence tensors for the instantiated CBW system.

\subsubsection{Watershed Engineering System Net Measurement Function} \label{subsec:cbw-watershedMeasurmentFunction}
The measurement of exogenously defined variables is largely dependent on the instantiated system. Subsection~\ref{subsec:cbw-CASTmeasurmentFunction} outlines the development of exogenously defined flows for the instantiated CBW.  

\subsubsection{Initial Conditions}
The initial conditions constraint (Eq.~\ref{eq:cbw-WLSE-watershed-initCond}) defines the initial stock of nutrients in each buffer to be zero. Thus, the final value of $Q_B$ is the mass retained in each buffer over the simulated period (Section~\ref{subsubsec:cbw-actwatershed}).

\section{Application to the Chesapeake Bay Watershed}\label{sec:cbw-applicationCBW}

This section describes the practical implementation of the water-specific HFGT–MBSE framework introduced in Section~\ref{sec:cbw-methods}. Subsection~\ref{subsec:cbw-CASTtransitionFunction} details the synthesis of the hetero-functional incidence matrix used to construct the Watershed Engineering System Net State Transition Function (Eq.~\ref{eq:cbw-WLSE-watershed-st}, Subsection~\ref{subsec:cbw-watershedTransitionFunction}). Subsection~\ref{subsec:cbw-CASTmeasurmentFunction} then explains how CAST source and scenario data inform the watershed engineering system net measurement function (Eq.~\ref{eq:cbw-WLSE-watershed-exogenous}, Subsection~\ref{subsec:cbw-watershedMeasurmentFunction}).

\subsection{Processing CAST Geospatial Data for the Engineering System Net State Transition Function} \label{subsec:cbw-CASTtransitionFunction}

The synthesis of the hetero-functional incidence tensor for the watershed engineering system net state transition function (Eq.~\ref{eq:cbw-WLSE-watershed-st}) begins with constructing a spatially explicit formal representation of the CBW. The purpose of this geospatial workflow is to generate the GIS-based resource definitions needed for the HFGT toolbox input file and, ultimately, the hetero-functional incidence matrix that encodes which capabilities inject or eject each operand for every CAST-defined buffer (e.g., “Land Segment X, nitrogen” or “Outlet Y, phosphorus”). These resources and capabilities follow directly from the watershed-specific BDD (Fig.~\ref{fig:cbw-bdd-watershed}) and its associated activity diagram 
(Fig.~\ref{fig:cbw-act-watershed}).

\begin{figure*}[htb!]
    \begin{subfigure}{0.19\linewidth}
        \centering
        \includegraphics[width=\linewidth]{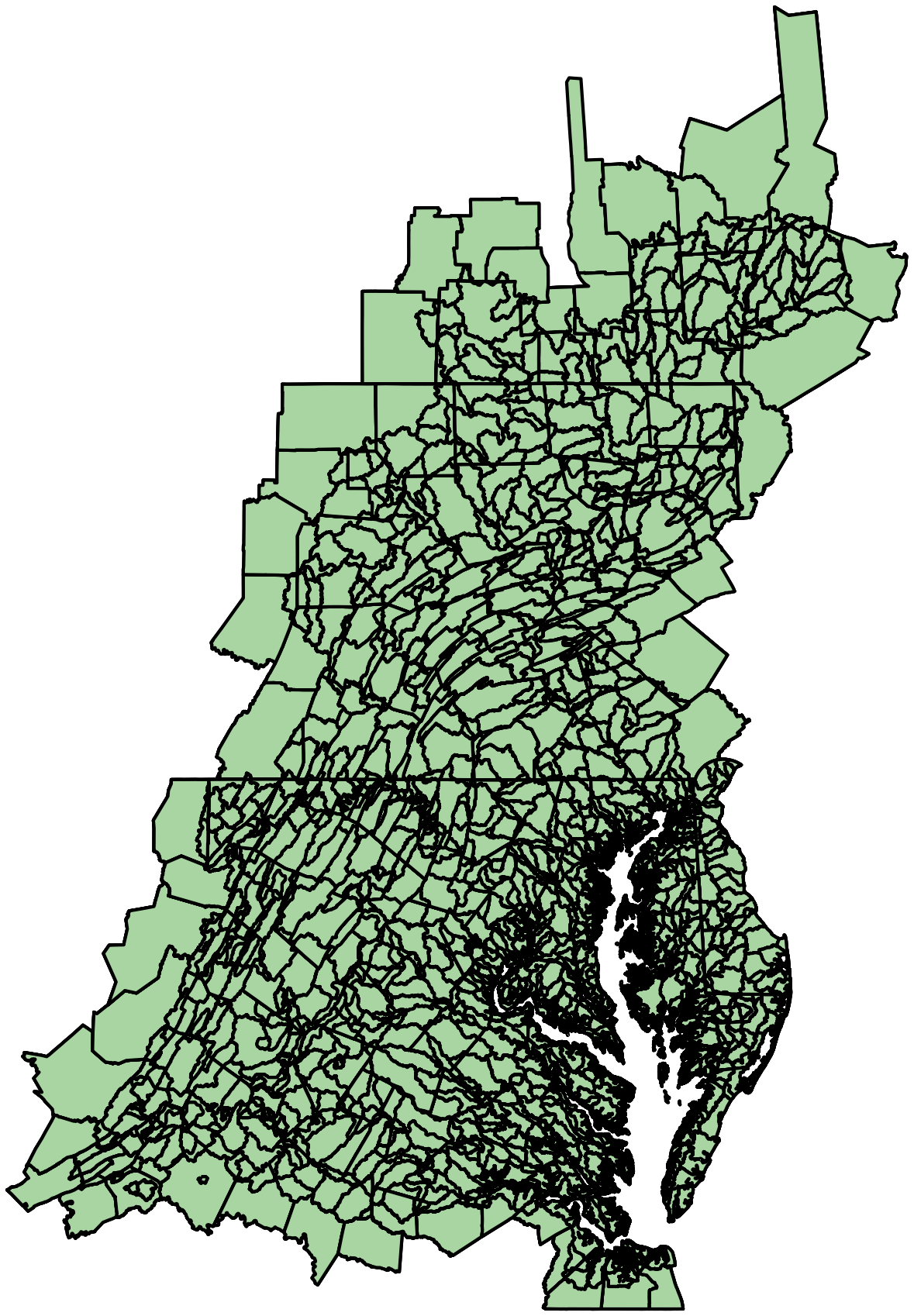}
        \caption{Land segments}
        \label{fig:cbw-CBlrsegments}
    \end{subfigure}
    \begin{subfigure}{0.19\linewidth}
        \centering
        \includegraphics[width=\linewidth]{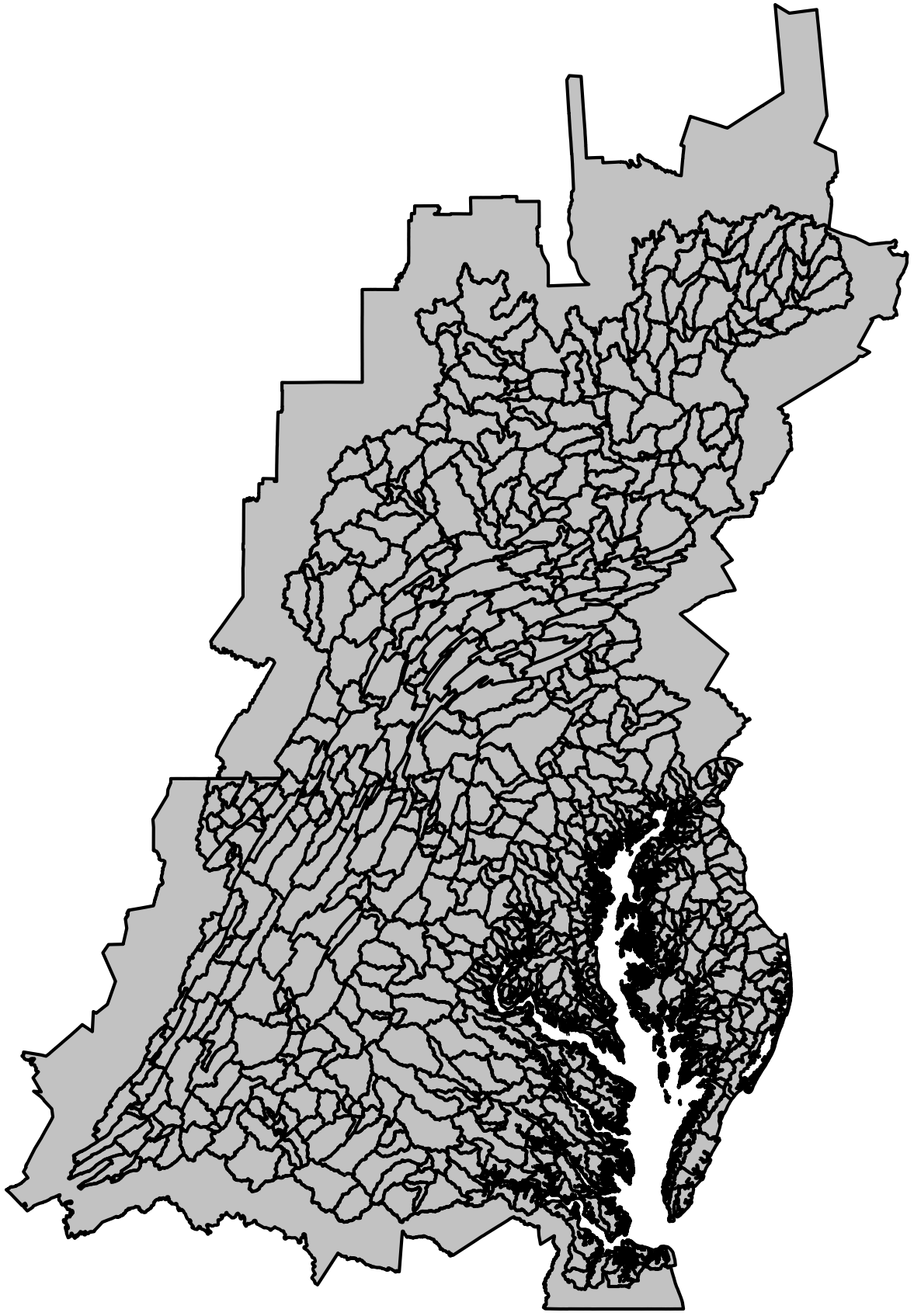}
        \caption{Subwatersheds}
        \label{fig:cbw-CBriversegs}
    \end{subfigure}
    \begin{subfigure}{0.19\linewidth}
        \centering
        \includegraphics[width=\linewidth]{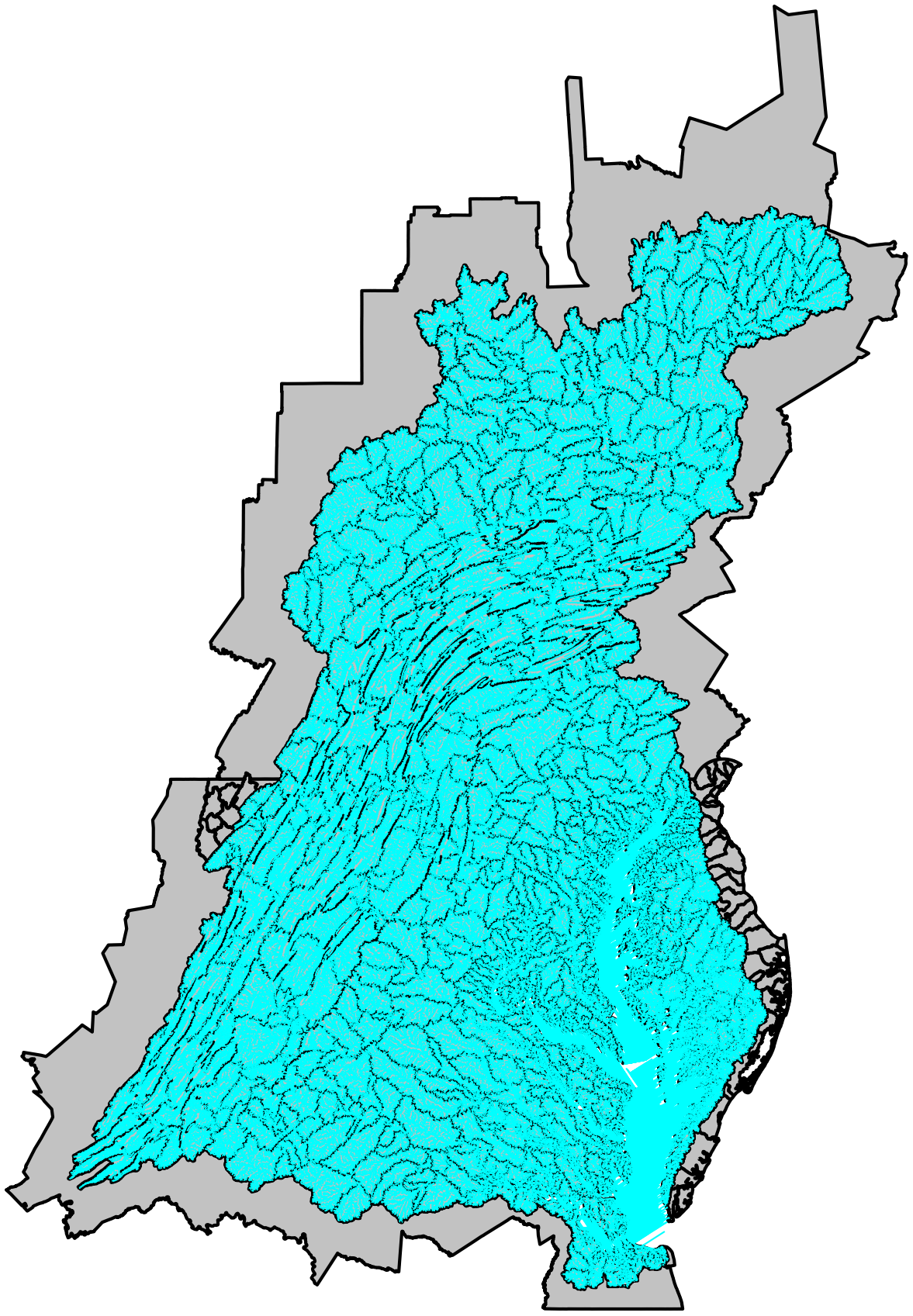}
        \caption{Streams}
        \label{fig:cbw-CBstreams}
    \end{subfigure}
    \begin{subfigure}{0.19\linewidth}
        \centering
        \includegraphics[width=\linewidth]{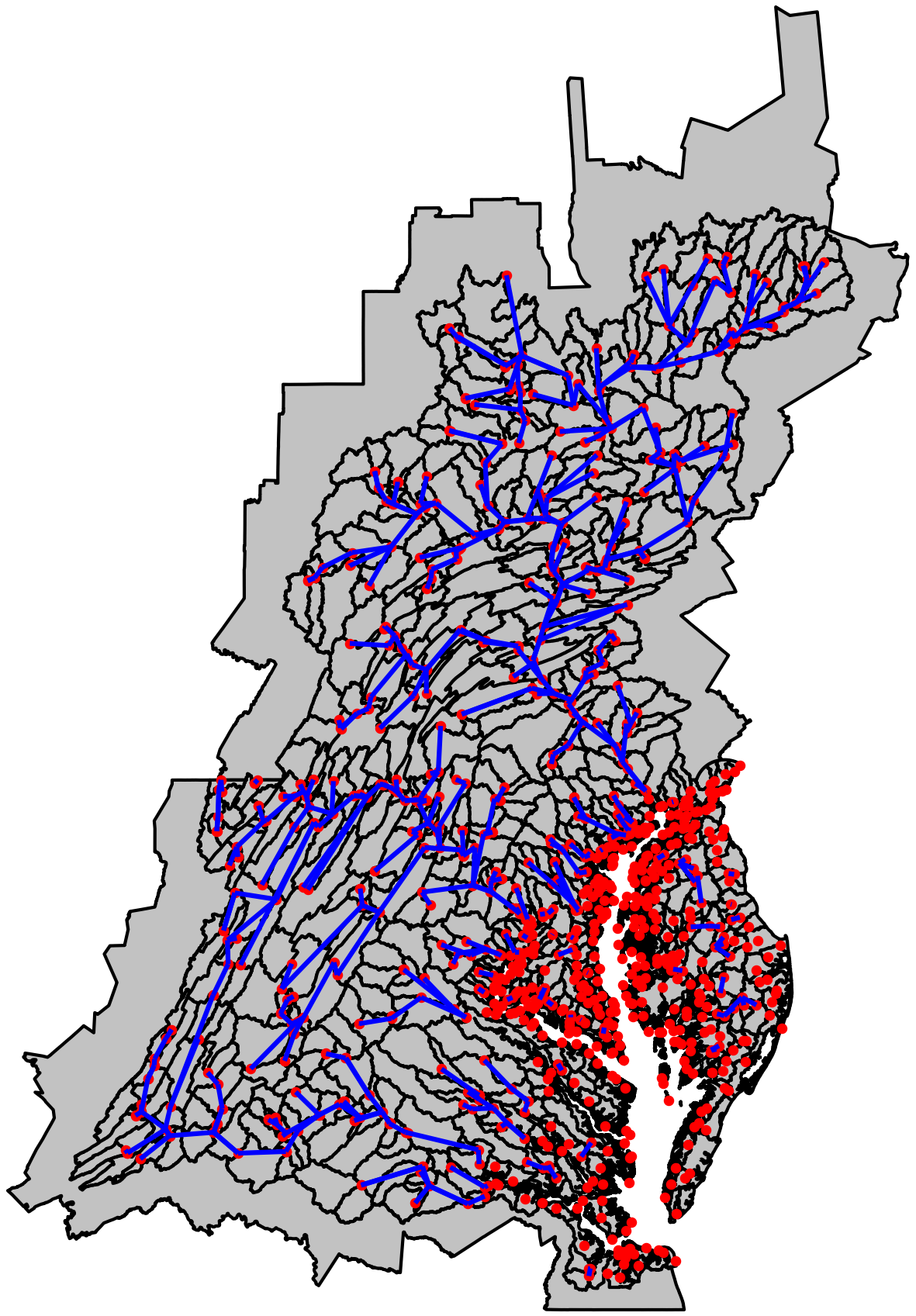}
        \caption{River Segments}
        \label{fig:cbw-CBnetwork}
    \end{subfigure}
    \begin{subfigure}{0.19\linewidth}
        \centering
        \includegraphics[width=\linewidth]{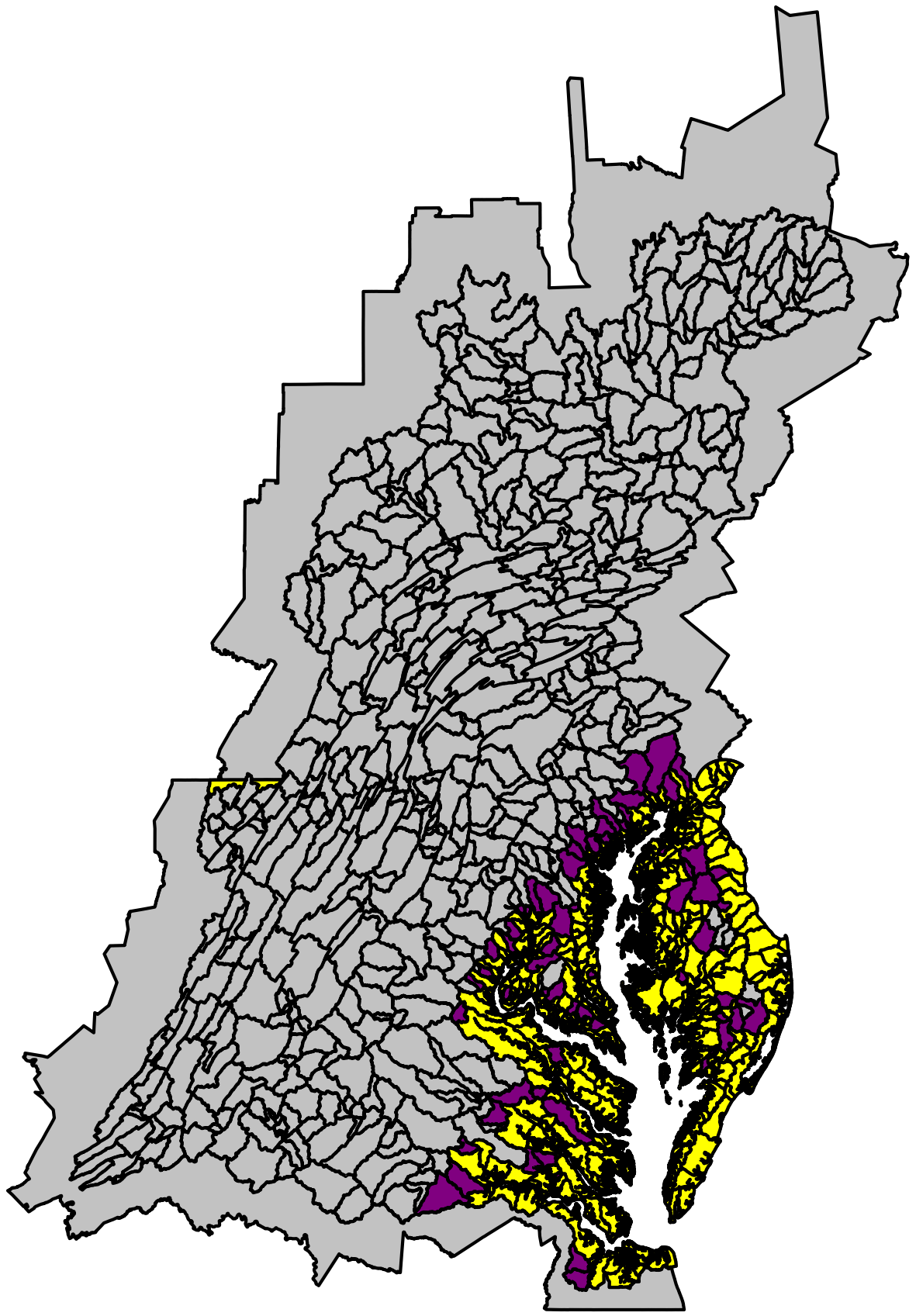}
        \caption{Estuary Links}
        \label{fig:cbw-CBestuarysegs}
    \end{subfigure}
    \caption{Derivation of a formal representation of the Chesapeake Bay watershed}
    \label{fig:cbw-chesapeakeBayMaps}
\end{figure*}

A spatially explicit model of the CBW was developed to reconstruct the implicit routing logic of CAST using publicly available geospatial data (Fig.~\ref{fig:cbw-chesapeakeBayMaps}). This representation defines how flows move from land–river segments through the stream network to the estuary, providing the structural foundation necessary for integrating watershed systems into the HFGT framework.

The workflow begins with CAST land–river segment polygons directly downloaded as a shapefile from the Chesapeake Bay Program (Fig.~\ref{fig:cbw-CBlrsegments}). These polygons are merged by shared subwatershed boundaries to generate river-segment polygons (Fig.~\ref{fig:cbw-CBriversegs}). Because the land-river segment polygons include counties that are only partially contained within the Chesapeake Bay watershed, some river segments drain to watersheds outside the Chesapeake Bay system. As a result, only a portion of the nutrient-generating activities within these counties contributes to nutrient transport to the Chesapeake Bay. 
These polygons occur along the perimeter of the Chesapeake Bay watershed (Fig.~\ref{fig:cbw-chesapeakeBayMaps}).

From there, constructing a spatial representation of downstream nutrient transport within the Chesapeake Bay watershed is complicated by the structure of the CAST land-river segments. The subwatershed component of each land-river segment is based on Hydrologic Unit Codes (HUCs) of varying orders rather than a uniform watershed delineation. This design choice was made by the Chesapeake Bay Program to produce land-river segments of comparable size across the watershed. While CAST identifies the downstream receiving river segment through its naming convention, it does not provide the geographic outlet locations needed to explicitly map connections among segments. Consequently, an additional procedure is required to estimate where flow exits each subwatershed and enters the downstream river network.

To approximate these outlet locations, a 30-m digital elevation model (DEM) was used to perform a flow-accumulation analysis for each subwatershed. Flow accumulation was calculated from the DEM, and the cell with the greatest accumulated upstream area within each segment was identified as a provisional outlet point. These outlet points provide an initial estimate of where water exits each subwatershed and serve as the basis for constructing the river network. Subsequent processing steps refine these locations and establish connectivity among river segments to represent downstream transport pathways throughout the watershed.


These provisional outlets are compared with the locations where CAST stream lines obtained from the Chesapeake Bay Program intersect each polygon (Fig.~\ref{fig:cbw-CBstreams}). Segment boundaries are converted to lines and intersected with the stream network, producing a dense set of candidate crossing points. Because the CAST stream dataset includes streams of multiple orders, there are substantially more intersection points than subwatersheds.

To determine the final outlet for each segment, provisional outlets are snapped to the most appropriate intersection point. Terminal segments are snapped to intersections on the estuary boundary, whereas internal segments are snapped to intersections consistent with downstream neighbors. This step aligns the spatial model with CAST naming conventions including trailing identifiers (e.g., “0000” for estuarine segments, “0001” for directly upstream segments).

After finalizing outlet locations, directed flowlines are constructed between upstream and downstream outlets (Fig.~\ref{fig:cbw-CBnetwork}). These relationships are inferred from the CAST segment naming structure, which encodes upstream and downstream connectivity. Segments draining directly to the estuary are linked to the nearest estuarine boundary (Fig.~\ref{fig:cbw-CBestuarysegs}). Spatial attributes are then harmonized. Segment identifiers and labels are updated for consistency. Centroids are then computed for visualization and aggregation, and outlet points are annotated with coordinates and flow direction metadata. Together, these operations produce a coherent and reproducible graph of watershed connectivity across the entire CBW.

With the geospatial representation complete, the resulting shapefiles are used to construct the HFGT toolbox input file that defines each resource type, operand, and capability according to the watershed RA (Fig.~\ref{fig:cbw-bdd-watershed} and Fig.~\ref{fig:cbw-act-watershed}).  For each buffer–operand pair (e.g., “Land Segment X, nitrogen”), the toolbox records which capabilities inject or eject each operand, as specified by the watershed activity diagram (Fig.~\ref{fig:cbw-act-watershed}). These relationships populate the hetero-functional incidence matrix, $M$, a core structural component of the HFGT formulation. This matrix, derived directly from the geospatial reconstruction, provides the structural basis for the watershed engineering system net state transition function used in the WLSEHFGSE formulation.

\subsection{Processing CAST Behavioral Data for the Measurement Function} \label{subsec:cbw-CASTmeasurmentFunction}


Complementing the hetero-functional incidence matrix described in the previous subsection, CAST behavioral data are used to derive the watershed engineering system net measurement function. Equation~\ref{eq:cbw-WLSE-watershed-exogenous} imposes exogenous measurements on the firing vector $U$, which represents all operand flows in the watershed. The selection matrix $D_U$ maps CAST-provided measurements to the corresponding capabilities in $U$, while the vector $C_U[k]$ contains time-step-specific estimates of operand flows derived directly from CAST datasets (Table~\ref{tab:cbw-cast-datasets}), with each entry corresponding to a specific watershed process or resource. For these routing equations the delivery factor enters $D_U$, and the corresponding entry of $C_U$ is zero.


The WLSEHFGSE formulation supports coarse or incomplete data in both space and time~\citep{Thompson:2025:ISC-JR11}. Rather than using a single selection matrix $D_U$, capability and temporal aggregation matrices--$D_{\mathcal{E}}$ (Defn.~\ref{defn:cbw-capabilityagmatrix}) and $D_T$ (Defn.~\ref{defn:cbw-temporalagmatrix})--aggregate firing-vector data temporally and spatially. These matrices are synthesized systematically using matrix algebra. For the present single-time-step implementation, $D_T$ reduces to the identity matrix, but additional time steps can be incorporated directly using CAST scenario data from multiple years.

\begin{defn}[Capability Aggregation Matrix $D_{\cal E}$~\citep{Thompson:2025:ISC-JR11}]\label{defn:cbw-capabilityagmatrix} The capability aggregation matrix $D_{\cal E} \in \{0,1\}^{|{\cal E}_S|\times |{\cal E}_D|}$ is a matrix whose element $D_{\cal E}(\psi_1,\psi_2)=1$ when the flow associated with system capability ${\epsilon}_{\psi_2} \in {\cal E}_S$ is part of the flow associated with data element $\widetilde{C}_U(\psi_1,k)$ at time k. 
\end{defn} 

\begin{defn}[Temporal Aggregation Matrix $D_{T}$~\citep{Thompson:2025:ISC-JR11}]\label{defn:cbw-temporalagmatrix} The temporal aggregation matrix $D_{T} \in \{0,1\}^{K\times K_D}$ is a matrix whose element $D_{T}(k_1,k_2)=1$ when the flows associated with time step $k_1 \in \{1, \ldots, K\}$ corresponds to the flows associated with $k_2 \in \{1, \ldots, K_D\}$ in the data $\widetilde{C}_U$. 
\end{defn}


This implementation represents an initial instantiation of the Chesapeake Bay watershed modeling suite within the HFGT Meta-architecture. The reference architecture includes two nutrient sources, agricultural and developed land, so every CAST quantity used below is restricted to these two sectors. Other CAST sources, such as point sources, septic systems, and atmospheric deposition to water, would require new source capabilities in the reference architecture and are left to future work (Section~\ref{sec:cbw-conclusion}).
\begin{table*}[!htbp]
    \centering
    \footnotesize
    \setlength{\tabcolsep}{4pt}
    \caption{CAST datasets used in the measurement function (CAST Phase~6, 2024 Progress scenario). Only the agricultural and developed sectors are used. EOS: edge of stream; EOR: edge of river; EOT: end of tide (delivered to the Bay).}
\begin{tabular}{
    >{\centering\arraybackslash}p{0.12\linewidth}
    >{\centering\arraybackslash}p{0.17\linewidth}
    >{\centering\arraybackslash}p{0.11\linewidth}
    >{\centering\arraybackslash}p{0.11\linewidth}
    >{\centering\arraybackslash}p{0.21\linewidth}
    >{\centering\arraybackslash}p{0.15\linewidth}
}
\toprule
\textbf{Dataset} & \textbf{Extracted data} & \textbf{Time} & \textbf{Spatial} & \textbf{Used for} & \textbf{Role} \\
\midrule
Nutrients Applied & Total applied N and P by sector & Annual (2024) & County & Accept (ag./dev.) (N/P) & Measurement \\
\midrule
Loads per Unit & EOS loading rate by load source and agency & Annual (2024) & County & Land segment EOS, then EOR & Builds measurement \\
\midrule
Base Conditions Report & Post-BMP acres by load source and agency & Annual (2024) & Land-river segment & Land segment EOS; weights for factors & Builds measurement \\
\midrule
Source Data Report & Stream-to-river and river-to-bay factors by load source & Steady state & Land-river segment & Land$\to$outlet (EOR); outlet$\to$outlet and outlet$\to$estuary routing & Builds measurement \\
\midrule
Loads Report & EOS by load source & Annual (2024) & County & Riparian pasture deposition; check of land segment EOS & Builds measurement; check \\
 & EOT by load source & Annual (2024) & County & Delivery to the Bay & Held out (test) \\
\bottomrule
\end{tabular}
    \label{tab:cbw-cast-datasets}
\end{table*}
Table~\ref{tab:cbw-cast-datasets} summarizes the CAST datasets that inform the engineering system net measurement function. Each measurement enters Eq.~\ref{eq:cbw-WLSE-watershed-exogenous} as one equation with its own error term in ${\cal E}_U$, so no measurement, including an aggregate total, is imposed exactly; the estimator balances all measurements against one another subject to exact mass balance (Eq.~\ref{eq:cbw-WLSE-watershed-st}).

\noindent \textbf{Accept (agricultural/developed) (nitrogen/phosphorus).}
The \textit{Nutrients Applied} dataset provides the measurements for all accept processes ($p = 1\text{--}4$) that introduce nitrogen and phosphorus into land buffers. For every county, sector, and nutrient, the sum of the corresponding accept firings in $U[k]$ over the county's land segments minus the CAST-reported total applied mass equals one error term in ${\cal E}_U$.

\noindent \textbf{Transport from land segments to outlet points (edge of river).}
In the reference architecture, a land segment's transport capability carries nutrients to its outlet point on the river network, which corresponds to CAST's edge-of-river (EOR) load. CAST reports EOS loads only by county, so EOS loads for each land-river segment were constructed from two CAST reports. For each load source and agency, the post-BMP acres in the segment (\textit{Base Conditions Report}) were multiplied by the county EOS loading rate for that load source and agency (\textit{Loads per Unit} report). Riparian pasture deposition, which has a load but no acreage, was allocated from the county total in the \textit{Loads Report} to segments in proportion to their pasture acres. This construction assumes one loading rate per load source and agency within a county. Loads are reported in million pounds per year (M~lb), the unit used by CAST; 1~M~lb $=$ 0.4536~Gg. Summed by county, the segment EOS loads reproduce CAST's county EOS loads for the two sectors to within 0.1\% (nitrogen 237.6 vs.\ 237.8~M~lb; phosphorus 11.25 vs.\ 11.26~M~lb; Supplementary Section~\ref{sec:cbw-SI-lrs-eos}). Each segment's EOS load by load source was then multiplied by its stream-to-river factor (\textit{Source Data Report}) and summed to give an EOR load for each of the 1,913 land-river segments (totals: nitrogen 199.9~M~lb, phosphorus 7.79~M~lb). Each land-to-outlet firing in $U[k]$ minus its segment's EOR load equals one error term. Because the county EOS loads were used to build these segment-level loads, they are not used again as separate measurements.

\noindent \textbf{Transport between outlet points and to the estuary.}
For each river segment, a river-to-bay factor was computed as the area-weighted average, over its land-river segments and load sources, of CAST's river-to-bay factors, using acres from the \textit{Base Conditions Report}. The fraction of the inflow to an outlet point that is passed to the next downstream outlet point is the ratio of the upstream to the downstream river-to-bay factor, and the fraction passed from an outlet point to the estuary is the river-to-bay factor of that terminal segment. Along any path to the Bay these ratios multiply to the river-to-bay factor of the starting segment, so if every land segment delivered exactly its EOR load, the network would deliver CAST's EOT load. For each such capability, the firing in $U[k]$ minus the delivery factor times the total inflow to its origin outlet point (the sum of the firings that inject into that buffer) equals one error term. River segments that contain no land area have no CAST factor; their 106 links pass flow through unchanged (factor of one). On 87 nitrogen links and 27 phosphorus links the ratio exceeds one (maximum 1.54 for nitrogen and 3.19 for phosphorus) because the downstream river-to-bay factor in CAST is smaller than the upstream one, i.e., CAST implies an in-river gain. Because outlet stocks are non-negative, a mass-conserving network cannot pass on more than it receives, so these measurements cannot be met exactly; their effect is quantified in Section~\ref{subsec:modelperformance}.

\noindent \textbf{Delivery to the Bay (end of tide), held out.}
CAST's EOT loads for the two sectors sum to 155.64~M~lb (70.6~Gg) of nitrogen and 5.694~M~lb (2.58~Gg) of phosphorus.
These totals could be added as one more measurement on the sum of all outlet-to-estuary firings. Instead, they are withheld from estimation and compared with the estimated delivery afterwards, as a \textit{hold-out test}: a quantity that is available is deliberately left out of estimation and then used to check the result, so agreement cannot come from fitting it directly. The EOT totals are withheld for two reasons. First, a single watershed total only weakly constrains how delivery is split among basins: when it is fitted, the estimator can match the total by shifting load between basins (Section~\ref{subsec:modelperformance}). Second, withholding it gives a test of the network that the fit itself does not guarantee. This test is not an independent validation against observations, because CAST computes EOT from the same factors used here; it checks whether the reconstructed network and estimator preserve CAST's delivery when CAST's loads and factors are imposed only as uncertain measurements.


\section{Results \& Discussion} \label{sec:cbw-results}
For this analysis, the reconstructed HFGT-based watershed model was executed for the ``2024 Progress'' scenario~\citep{chesapeake-bay-program:2024:00} as a single annual time step. While the watershed reference architecture and WLSEHFGSE formulation support multiple time steps, this study focuses on model development and demonstration. All results are from the primary configuration, in which CAST's EOT totals are withheld from estimation (Section~\ref{subsec:cbw-CASTmeasurmentFunction}); a second run in which the EOT totals are fitted is used only for comparison in Section~\ref{subsec:modelperformance}.

Subsection \ref{subsec:cbw-modeloutputs} summarizes the reconstructed nitrogen and phosphorus outputs of this simulation run. Subsection \ref{subsec:modelperformance} evaluates model performance, and Subsection \ref{subsec:implications} discusses implications. 

\subsection{Model Outputs}\label{subsec:cbw-modeloutputs}

\begin{figure*}[htb!]
    \centering
    \includegraphics[width=\linewidth]{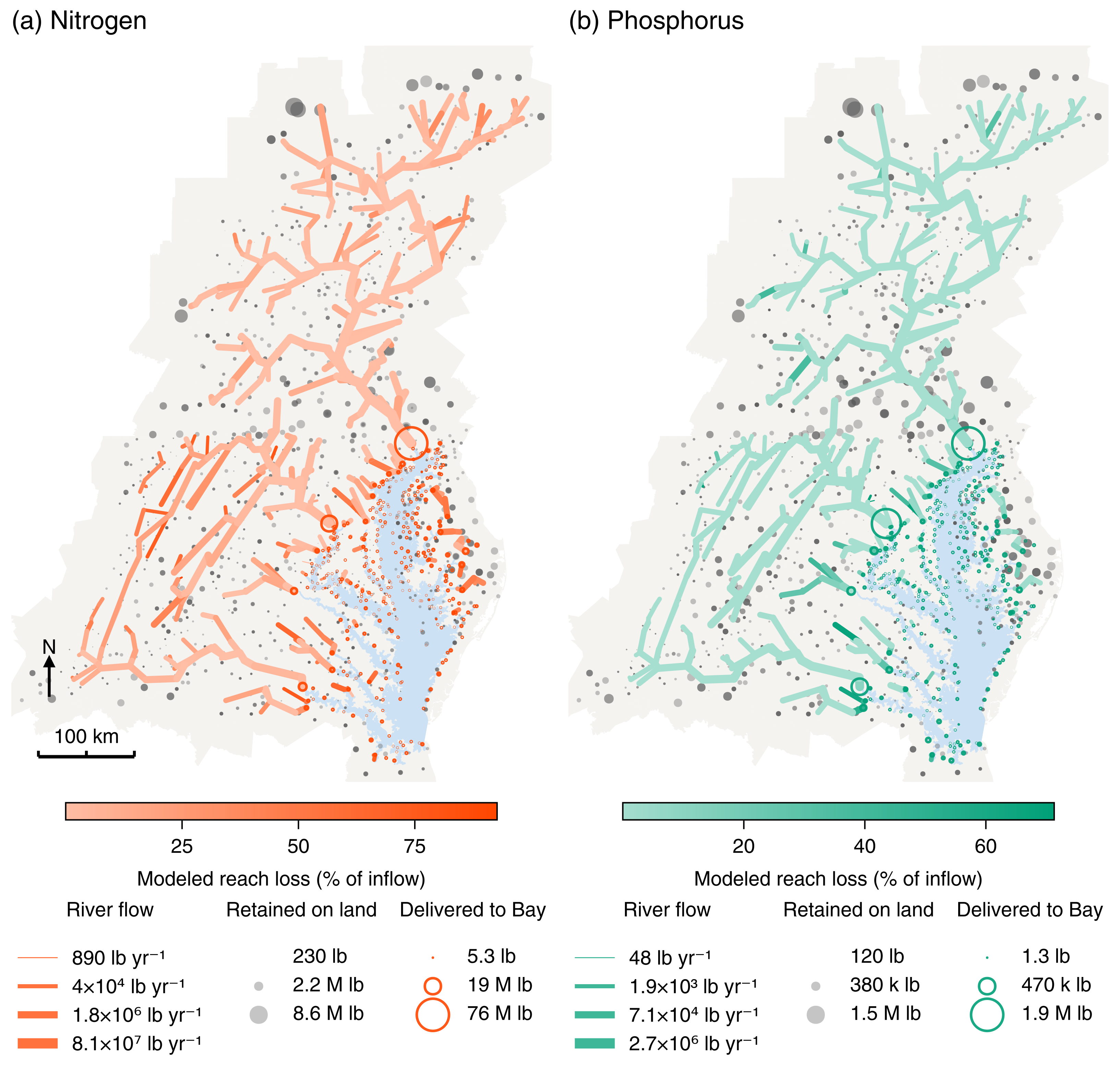}
    \caption{Estimated (a) nitrogen and (b) phosphorus transport on the reconstructed river network, 2024 Progress scenario, with CAST's delivery to the Bay withheld from estimation. Line width shows the estimated flow between outlet points (log scale, from the smallest to the largest drawn flow); flows below 1~lb~yr$^{-1}$ are not drawn. Line color shows the modeled loss on each reach, the outlet's retained mass as a percentage of its inflow. Gray dots show the mass retained on each land segment (applied minus delivered to the river network). Open circles show the load delivered to the Bay at each outlet point that drains directly to the estuary.}
    \label{fig:cbw-map-network}
\end{figure*}

Figure~\ref{fig:cbw-map-network} shows the estimated state of the network. Most of the applied mass is retained on land (gray dots; 736~M~lb of nitrogen and 179~M~lb of phosphorus summed over land segments), consistent with CAST's land-to-water and stream-to-river factors, which are built into the EOR measurements. In the HFGT representation, land retention is quantified within the buffer storage $Q_B$ of each land segment: the applied mass that is not transported to the outlet point. Because the reference architecture has no separate capabilities for crop removal, storage in soils, or other losses, this single value combines all of them~(Section~\ref{subsubsec:cbw-actwatershed}).
 Land retention represents the difference in nutrient inflow and outflow within each land segment. Of the mass that reaches the river network, 143.2~M~lb of nitrogen and 5.37~M~lb of phosphorus are delivered to the Bay through 422 terminal outlet points (open circles), with the largest deliveries from the Susquehanna and Potomac outlets.

The line colors show where mass is lost across the river network. Although CAST assigns a river-to-bay factor to each land-river segment, it does not identify the individual reaches where the corresponding losses occur. The network reconstruction makes this spatial pattern explicit. For nitrogen, losses are distributed across most reaches, with approximately zero loss on only 73 of the 587 drawn reaches. For phosphorus, in contrast, 498 of the 595 drawn reaches have approximately zero loss, concentrating attenuation on roughly one hundred reaches. This pattern reflects CAST's fitted factors: the phosphorus river-to-bay factor is identical for consecutive river segments on 504 of 610 reaches. These results describe the spatial distribution of attenuation implied by CAST's parameterization, not observed patterns of physical phosphorus retention. Nevertheless, they reveal a structural property of CAST that is not apparent without an explicit network representation.

The estimated relative contribution of the agricultural and developed sectors to applied nutrients on each land segment is mapped in Supplementary Fig.~\ref{fig:cbw-map-sectorShar}.

\subsection{Model Performance}\label{subsec:modelperformance}
Table~\ref{tab:cbw-wlse-fit-summary} compares two runs: the primary run, in which CAST's EOT totals are withheld from estimation, and a comparison run in which they are fitted as measurements. CAST-derived quantities enter the WLSEHFGSE as measurements rather than exact constraints, and they are not mutually consistent everywhere, so exact reproduction of every measurement is not expected.

\begin{table*}[!htbp]
\centering
\small\linespread{1}\selectfont
\caption{Fit to the measurements and delivery to the Bay with CAST's EOT withheld from estimation (primary run) and with EOT fitted as a measurement (comparison run), 2024 Progress scenario, agricultural and developed sectors. R\textsuperscript{2} and NRMSE (RMSE divided by the mean measured value) are computed across county--sector pairs (applied) and across counties (land$\to$outlet).}
\begin{tabular}{>{\raggedright\arraybackslash}p{0.25\linewidth} >{\raggedright\arraybackslash}p{0.13\linewidth} *{4}{>{\centering\arraybackslash}p{0.11\linewidth}}}
\toprule
 & & \multicolumn{2}{c}{\textbf{EOT held out}} & \multicolumn{2}{c}{\textbf{EOT fitted}} \\
\cmidrule(lr){3-4} \cmidrule(lr){5-6}
\textbf{Quantity} & \textbf{Metric} & \textbf{N} & \textbf{P} & \textbf{N} & \textbf{P} \\
\midrule
Applied, by county and sector & R\textsuperscript{2} & 1.000 & 1.000 & 1.000 & 1.000 \\
 & NRMSE & 0.16\% & 0.05\% & 0.20\% & 0.05\% \\
Land$\to$outlet vs.\ CAST-derived EOR & R\textsuperscript{2} & 0.997 & 0.997 & 0.995 & 0.996 \\
 & NRMSE & 8.9\% & 8.6\% & 11.0\% & 10.0\% \\
\midrule
Delivery to the Bay (M lb; CAST EOT: 155.64 N, 5.694 P) & Estimated & 143.22 & 5.368 & 155.63 & 5.694 \\
 & Difference & $-8.0\%$ & $-5.7\%$ & $-0.01\%$ & $<0.01\%$ \\
\bottomrule
\end{tabular}
\label{tab:cbw-wlse-fit-summary}
\end{table*}

In both runs the applied loads are reproduced almost exactly ($R^2=1.000$). The estimated land-to-outlet transport agrees closely with the CAST-derived EOR loads when summed by county ($R^2=0.997$, NRMSE 8.9\% for nitrogen and 8.6\% for phosphorus), and this agreement worsens when the EOT totals are fitted (NRMSE 11.0\% and 10.0\%), because matching the watershed total requires the land-to-outlet flows to depart from their measurements.

When the EOT totals are fitted, the estimator matches them almost exactly, but it does so by moving delivery among basins: relative to routing CAST's EOR loads with CAST's factors, nitrogen delivery from the Potomac falls from 27.0 to 19.0~M~lb while delivery from the Susquehanna rises from 75.9 to 79.7~M~lb and from the James from 4.2 to 5.4~M~lb. Agreement with a single watershed total is therefore not evidence that the network routes load correctly. In the primary run, which never sees the EOT totals, the estimated delivery is 8.0\% below CAST's EOT for nitrogen and 5.7\% below for phosphorus.

This difference has identifiable causes. Routing CAST's EOR loads through the network with CAST's own factors, with no estimation, reproduces CAST's EOT to  within 0.1\% for nitrogen and 0.3\% for phosphorus, so the reconstructed network preserves CAST's delivery structure. Capping the delivery factors greater than one at one, as mass conservation requires, accounts for 2.7 percentage points of the nitrogen difference and 1.8 of the phosphorus difference. The remainder arises because the estimator cannot satisfy all measurements at once: for example, 139 land segments have nitrogen EOR loads larger than the nitrogen applied to them (a total of 1.18~M~lb), most likely because CAST's developed-land load includes atmospheric deposition, which is not part of the applied data.

\subsection{Implications}\label{subsec:implications}

Re-expressing CAST as an explicit, mass-conserving network reveals two features of its statistical fitting that are not apparent from its segment-to-Bay delivery factors alone. First, it localizes the implied mass losses along the river network, showing that phosphorus attenuation is concentrated on a small fraction of reaches. Second, it identifies fitted values that are inconsistent with mass conservation: delivery factors greater than one on 87 nitrogen and 27 phosphorus links, and land segments whose EOR loads exceed their applied loads. These inconsistencies arise because CAST's statistical fitting does not enforce mass balance along the network. Together, these findings identify where an all-source reference architecture or refitted delivery factors could most substantially change the reconstruction.

Several limitations qualify these findings. Because all measurements are weighted equally, the objective function gives greater influence to large flows. The reconstruction represents only the agricultural and developed sectors, accounting for 155.6~M~lb of nitrogen and 5.69~M~lb of phosphorus delivered to the Bay in CAST, rather than the total Bay load. The hold-out test evaluates consistency with CAST, not with observations; the reconstruction has not yet been validated against monitored loads. Finally, the model assumes that nutrient transport follows the dendritic river network derived from watershed topology. It does not explicitly represent processes that alter natural routing, including reservoirs, groundwater transport, pumping infrastructure, and interbasin transfers.
Beyond reconstructing watershed connectivity, the explicit network representation supports dynamic simulation, data assimilation, and integration with infrastructure and governance models. Because each data source enters with its own error term, monitored loads can later be added as further measurements on the same network, and the estimator will then weigh CAST's fitted factors against observations rather than against each other.

\section{Conclusion} \label{sec:cbw-conclusion}

This study develops an HFGT-based structural reconstruction of a real-world watershed system, demonstrating how MBSE and HFGT can formalize the physical and functional structure underlying an existing empirical model. Using CAST applied loads, land-segment edge-of-river loads, and river-to-bay delivery factors as measurements, each with its own error term, the estimator reproduces CAST's inputs closely and, with CAST's delivery to the Bay withheld, predicts that delivery to within 8.0\% for nitrogen and 5.7\% for phosphorus. The difference is explained by delivery factors greater than one, which mass-conserving routing cannot reproduce, and by conflicts among CAST's inputs that the estimator must reconcile.


The explicit network also shows where CAST's delivery implies loss along the river network, and that for phosphorus this loss is concentrated on a small fraction of reaches. These are properties of CAST's fitted factors that become visible only when the model is required to conserve mass along a network.
Future work will extend the reconstruction in four directions: (1) incorporate all of CAST's nutrient sources, including point sources, septic systems, and atmospheric deposition, to represent the total load to the Bay; (2) evaluate estimated flows against monitored loads; (3) weight measurements according to data provenance; and (4) couple the watershed with the estuary model and incorporate hydrologic variables from process-based watershed models. Together, these extensions will improve the representation and evaluation of nutrient transport while preserving the transparency of the systems architecture.

Beyond the watershed system, the extensibility of the MBSE-HFGT architecture enables integration with models of human systems that shape nutrient dynamics. Economic and governance models can represent decision-making, incentives, and institutional feedbacks that influence nutrient management across jurisdictions. Because the accept capabilities of the watershed reference architecture are defined by economic sector, they provide a natural coupling point to economic input--output models, which have already been represented in MBSE--HFGT with results identical to their traditional formulation~\citep{naderi:2026:01}. Embedding these processes within the shared architecture enables multi-domain scenario evaluation, a step toward the system-of-systems convergence paradigm proposed by Little et al.~\citep{Little:2026:ISC-JR10}.

The primary contribution of this work is a transferable methodology for representing watershed systems within the MBSE-HFGT architecture. By embedding an existing empirical watershed model in a transparent systems representation, the framework supports integration with environmental, infrastructure, economic, and governance models while preserving the scientific basis of the original model. The approach complements existing watershed models by providing a common systems representation for interdisciplinary analysis and decision support.

\section*{Acknowledgments}\label{sec:simplehydro-lakes-Acknowledgments}
This research is based on work supported by the Growing Convergence Research Program of the National Science Foundation under Grant Numbers OIA 2317874 and OIA 2317877. 

\section*{Data and Code Availability}

The modeling framework presented in this study was implemented using the Hetero-functional Graph Theory (HFGT) toolbox and custom Python and Julia scripts developed by the authors. The code required to reproduce the simulation results is publicly available in a version-controlled GitHub repository at \url{https://github.com/LIINES/HFGT-CBW-Watershed-Model}. The results in this paper were produced with release \texttt{v2.0} of the repository.

The data associated with the Chesapeake Bay watershed simulation, along with detailed instructions for constructing the required directory structure, are archived on Zenodo at \url{https://doi.org/10.5281/zenodo.18763339}. 




This study uses publicly available data products from the Chesapeake Assessment Scenario Tool (CAST), developed by the Chesapeake Bay Program \citep{chesapeake-bay-program:2024:00}, together with associated Chesapeake Bay Program geospatial datasets \citep{chesapeake-bay-program:2020:00}. All model inputs were derived from these public sources through documented preprocessing and aggregation steps.

No proprietary or restricted datasets were used in this study. Reproduction of the results requires a standard personal computer and the software listed in the repository. The quadratic program was solved with Gurobi under a free academic license.

\section*{CRediT Author Statement}
\textbf{Dr. Megan S. Harris:} Conceptualization, Methodology, Software, Validation, Formal analysis, Investigation, Writing - Original Draft, Visualization, Project Administration.
\textbf{Dr. John C. Little:} Conceptualization, Resources, Writing - Review \& Editing, Supervision, Project Administration, Funding Acquisition.				
\textbf{Dr. Amro M. Farid:}	Conceptualization, Methodology, Software, Validation, Writing - Review \& Editing, Supervision, Project Administration, Funding Acquisition.

\section*{Declaration of generative AI and AI-assisted technologies}
During the preparation of this work the authors used ChatGPT (OpenAI) and Claude (Anthropic) to assist with code development and manuscript editing. The authors reviewed and edited all code and text and take full responsibility for the content of the publication.
\section*{Appendix A. Supplementary data}
Supplementary material associated with this article (background on HFGT and the WLSEHFGSE, the sector-share map of applied nutrients, and the construction and check of the land-segment edge-of-stream loads) is provided in the online version.

\bibliographystyle{elsarticle-harv}
\bibliography{LIINESLibrary, LIINESPublications,
megBibliography
}

\includepdf[pages=-]{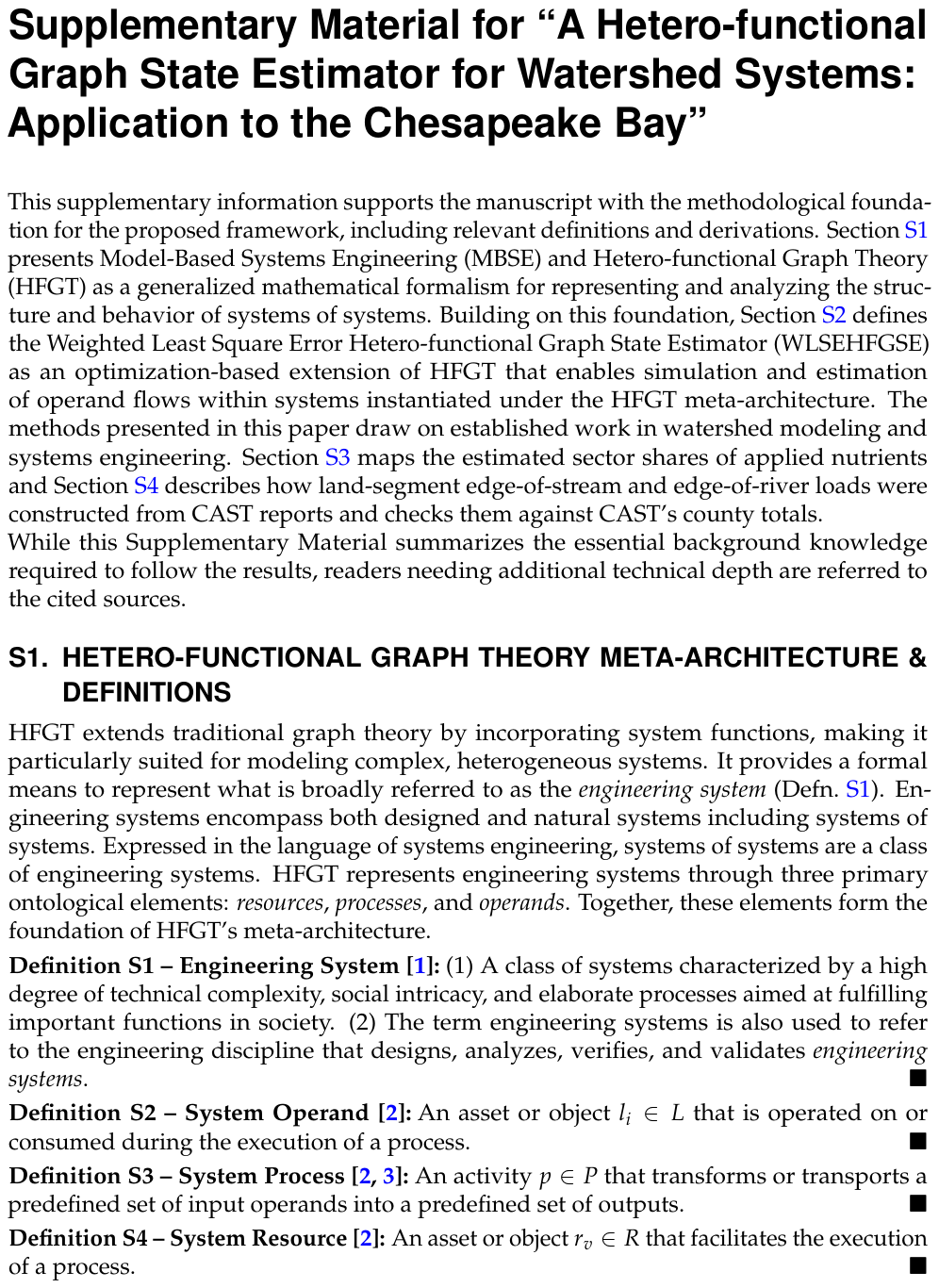}
\end{document}